\documentclass[%
 showkeys,
superscriptaddress,
nofootinbib,
 amsmath,amssymb,
 aps, physrev,
 twocolumn,
floatfix,
10pt
]{revtex4-2}

\usepackage{graphicx}
\usepackage{dcolumn}
\usepackage{bm}
\usepackage{physics}
\usepackage[thinc]{esdiff}
\usepackage[lofdepth,lotdepth]{subfig}
\usepackage{enumitem}

\begin{document}

\preprint{APS/123-QED}

\title{\textbf{Revisiting the propagation of highly-energetic gamma rays in the Galaxy} 
}%

\author{Gaetano Di Marco}
  \email{Contact author: gaetano.dimarco@ift.csic.es}
\affiliation{
 Instituto de F\'isica Te\'orica UAM/CSIC,
Calle Nicol\'as Cabrera 13-15, Cantoblanco, 28049 Madrid, Spain
}%
\affiliation{
 Departamento de Física Teórica, M-15, Universidad Autónoma de Madrid, E-28049 Madrid, Spain
}%

\author{Rafael \surname{Alves Batista}}%
 \email{rafael.alves$\_$batista@iap.fr}
\affiliation{
 Instituto de F\'isica Te\'orica UAM/CSIC,
Calle Nicol\'as Cabrera 13-15, Cantoblanco, 28049 Madrid, Spain
}%
\affiliation{
 Departamento de Física Teórica, M-15, Universidad Autónoma de Madrid, E-28049 Madrid, Spain
}%
\affiliation{Sorbonne Universit\'e, CNRS, UMR 7095, Institut d'Astrophysique de Paris, 98 bis bd Arago, 75014 Paris, France}%


\author{and Miguel A. S\'anchez-Conde}
 \email{miguel.sanchezconde@uam.es}
\affiliation{
 Instituto de F\'isica Te\'orica UAM/CSIC,
Calle Nicol\'as Cabrera 13-15, Cantoblanco, 28049 Madrid, Spain
}%
\affiliation{
 Departamento de Física Teórica, M-15, Universidad Autónoma de Madrid, E-28049 Madrid, Spain
}%

\date{\today}

\begin{abstract}
Recent gamma-ray observations have detected photons up to energies of a few PeV. These highly energetic gamma rays are emitted by the most powerful sources in the Galaxy. Propagating over astrophysical distances, gamma rays might interact with background photons producing electron-positron pairs, then deflected by astrophysical magnetic fields. In turn, these charged particles might scatter through inverse Compton galactic radiation fields, triggering electromagnetic cascades. In this scenario, the characterisation of astrophysical environment in which gamma rays travel, specifically background photons and magnetic fields, is crucial. We explore the impact of propagation effects on observables at Earth by simulating galactic sources emitting gamma rays with energies between $100 \; \text{GeV}$ and $100 \; \text{PeV}$. We analyse the imprint of the galactic environment on observed energy spectra and arrival direction maps, revealing gamma-ray absorption features in the former and ``deflection" of gamma rays in the latter. Specifically, owing to interstellar radiation field spatial distribution and the galactic magnetic field structure, propagation effects on observables are found to be related to the specific gamma-ray source position and to the prompt emission model. Detailed investigations of the propagation effect on galactic gamma rays will improve the robustness of both current and future gamma-ray detections and indirect dark matter searches.
\end{abstract}

\maketitle


\section{Introduction}

The past decade has been a fervent and vivid era for gamma-ray astronomy thanks to the activity of space- and ground-based experiments. Among the former are the Fermi-LAT (Large Area Telescope)~\cite{atwood2009large} and AGILE (\textit{Astro-Rilevatore Gamma a Immagini Leggero})~\cite{basset2007agile} detectors, amid the latter are MAGIC (Major Atmospheric Gamma Imaging Cherenkov)~\cite{magicSite} and H.~E.~S.~S. (High Energy Stereoscopic System)~\cite{HESSSite} telescopes, LHAASO (Large High Altitude Air Shower Observatory)~\cite{zhen2010future} and HAWC (High-Altitude Water Cherenkov)~\cite{westerhoff2012hawc} observatories. Moreover, the approaching dawn of CTAO (Cherenkov Telescope Array Observatory)~\cite{acharya2019science}, ASTRI (\textit{Astrofisica con
Specchi a Tecnologia Replicante Italiana}) Mini-Array~\cite{pareschi2016astri} and SWGO (Southern Wide-field Gamma-ray Observatory)~\cite{abreu2019southern} observatories will enable the exploration of the gamma-ray sky with unprecedented performances~\cite{celli24detprospects}. In this context, recent detection of galactic gamma rays with energies ranging from several hundreds of TeV up to few PeVs~\cite{cao2021ultrahigh, cao2023ultrahigh}, in some cases showing extended morphologies at the highest energy bands~\cite{albert2023discovery, cao2024does}, pose new theoretical challenges in the understanding of the sources. 

Emissions of gamma rays with energies beyond tens of TeV are likely produced by galactic PeVatrons, efficient accelerators of cosmic rays up to several PeVs~\cite{higdon2005ob,gabici2007searching,bykov2014nonthermal,albert2020hawc,cao2021ultrahigh,cardillo2023lhaaso,costa2023gamma, wilhelmi2024hunt}. Both hadronic and/or leptonic mechanisms have been proposed to explain the consequent gamma-ray emission from PeVatrons. In the former, gamma rays are generated from decays of neutral pions, secondaries of hadronic processes that involve cosmic rays accelerated by PeVatrons~\cite{kar2022ultrahigh, de2023detection, mitchell2023lhaaso}. Leptonic scenarios explain the production of gamma rays with energy larger than 100~TeV in young pulsars placed in radiation-dominated regions~\cite{vannoni2009diffusive,breuhaus2021ultra,breuhaus2022pulsar}.

In addition, fluxes of photons with energies larger than $10$~PeV are predicted from ultra-high-energy cosmic-ray interactions with cosmological photon fields, the so-called cosmogenic photons~\cite{murase2012b, berezinsky2016a, alvesbatista2019a}, and from primordial relics such as in top-down models~\cite{fodor2001grand, sarkar2002high, ellis2006ultrahigh}, thus super-heavy dark matter~\cite{berezinsky1997ultrahigh, birkel1998extremely, kalashev2016constraining, he2020expected, esmaili2021first, das2023revisiting, cao2024constraints}, Z-burst scenarios~\cite{weiler1982resonant, fargion1999ultra} or topological defects~\cite{hill1983monopolonium, hindmarsh1995cosmic}. Nevertheless, many of these models have been strongly constrained already~\cite{auger2017d, ta2019a, auger2022c,deligny2024searches}. In the highest energy bands, the upper limits on the gamma-ray diffuse flux above $\sim 10^{17} \; \text{eV}$ from the Pierre Auger Observatory have been employed~\cite{halim2024search}.

Apart from production mechanisms, gamma rays propagate over astrophysical distances before eventually reaching our observatories. During their travel, they might interact with background photons permeating astrophysical spaces, producing electron-positron pairs, to which from now on we refer simply as \textit{electrons}. Such electrons are then deflected by astrophysical magnetic fields and, in turn, they might scatter through inverse Compton (IC) background photons up to several TeVs. These processes trigger electromagnetic (EM) cascades over astrophysical distances\footnote{Throughout the text, even with a low number of occurring interactions we refer to the phenomena as EM cascade.}. In this picture, prompt gamma-ray fluxes could be distorted during their travel towards detectors on Earth. Specifically, their energy spectra could present absorption features and arrival direction maps with spreading counts due to ``deflected'' gamma rays. The latter might lead to the formation of haloes, i.e. extended gamma-ray counts surrounding the emission cores. Moreover, in this picture, the deflection of electrons in the EM cascade produce delays in the arrival times of gamma rays. These effects have been largely studied in the context of gamma rays propagating over intergalactic lengths~\cite{neronov2009a}, leading to constraints on intergalactic magnetic field properties~\cite{neronov2010a,alvesbatista2016b,ackermann2018search,alves2021gamma}. Propagation effects are fundamental not only in the understanding of source production mechanisms and intergalactic spaces, but also for peculiar indirect searches for axion-like particles~\cite{raffelt1988mixing,wouters2014anisotropy,troitsky2016towards,carenza2021turbulent,eckner2022first,zhang2023axion}, dark matter candidates. 

The goal of this paper is to explore the influence of propagation effects on gamma rays travelling over galactic length scales, after leaving their production region. To achieve this, it is crucial to characterise the galactic environment. Besides the cosmic microwave background (CMB) and cosmic radio background (CRB), galactic space is permeated by the interstellar radiation field (ISRF). In intergalactic space, the equivalent of the ISRF is the extragalactic background light (EBL)~\cite{franceschini2008extragalactic,finke2010modeling,dominguez2011extragalactic,gilmore2012semi,stecker2016empirical,saldana2021observational,finke2022modeling}. Previous works investigated gamma-ray galactic absorption due to pair production on ISRF and CMB by computing the optical depth~\cite{moskalenko2006attenuation,zhang2006very,esmaili2015gamma,vernetto2016absorption,popescu2017radiation,porter2018galactic}. They found that gamma-ray fluxes from sources located in the galactic center start to be attenuated at energies above $\sim~10$~TeV. Moreover, our Galaxy has a peculiar magnetic field structure, product of thermal electrons pervading interstellar space, large-scale plasma bubbles and small-scale turbulent features due to outflows, then enhanced by hydrodynamic turbulences. Galactic magnetic field (GMF) models are few and carry uncertainties, given the complexity and variety of the measurements involved~\cite{jansson2012new,adam2016planck,terral2017constraints,kleimann2019solenoidal}.

This work aims to investigate theoretical scenarios of gamma rays, with energy between 100~GeV~and~100~PeV, propagating in our Galaxy, using dedicated Monte Carlo (MC) simulation tools. Unlike the previously cited works, these tools also enable us to trace the three-dimensional development of EM cascades, reconstructing accordingly the galactic environment. It is organised as follows: in Sec.~\ref{simulationSetting} the simulation settings are described together with gamma-ray propagation theory and galactic environmental properties. In Sec.~\ref{result_sec}, the observables at Earth position from the simulated gamma-ray sources are examined. Chiefly, spectral and spatial observables are related to the specific source position in the Galaxy, because of background photons and magnetic field spatial models. Sec.~\ref{conclPersp} outlines the conclusions of this work and open questions to address in future follow-up studies.

\section{\label{simulationSetting} Simulation of galactic gamma-ray propagation}

Gamma rays propagating in our Galaxy are simulated within the CRPropa~3.2 framework~\cite{Batista_2016,batista2022crpropa}. Its modular structure enables the definition of observer, gamma-ray sources, astrophysical environmental properties, i.e. background photons and magnetic fields, interactions in play and equation of motion solvers. Useful functions allow the user to define simulation boundaries, both in space and energy. 

The Boris push algorithm~\cite{boris1972proceedings} is employed to solve the equations of motion of the charged particles moving in magnetic field regions. The minimum step size is set to a fraction of the smallest estimated Larmor radius of the propagating (charged) particles. It is computed as $R_{L}=E_{e}/(ecB)$, where $E_{e}$ is the electron energy, $B$ the magnetic field intensity, and $e$ and $c$ are the elementary electric charge and the speed of light, respectively. This choice permits deflections to be accurately computed. Secondary electrons have, on average, an energy $E_{e}=E_{\gamma_{0}}/2$, where $E_{\gamma_{0}}$ is the primary gamma-ray energy. The parameter estimating the amount of the primary particle energy channeled to the interaction products is the \textit{inelasticity}~\cite{esmaeili2024neutrinos}, that can be very large for EM processes. Within each simulation step, interaction probabilities are randomly drawn via the MC algorithms implemented in CRPropa, which could eventually generate the interaction products. 

\subsection{Interactions and background photon fields}\label{int_RD_subsec}

In general, the inverse mean free path of gamma ray or electron of energy $E$, moving through an isotropic background photon field with number density $n(\epsilon, z)$, is computed as~\cite{alvesbatista2016b}: 
\begin{equation}\label{InvMFP}
    \lambda^{-1}(E,z)=\frac{1}{8E^{2}}\int_{0}^{\infty
}\dd \epsilon\int_{s_{\text{min}}}^{s_{\text{max}}
}\dd s\frac{1}{\epsilon^{2}}\diff{n(\epsilon,z)}{\epsilon}\mathcal{F}(s) \,,
\end{equation}
where $z$ is the redshift, $\epsilon$ the background photon energy and $s$ the center-of-mass energy squared. The $\mathcal{F}(s)$ function is process-dependent, and so are the kinematic limits $s_{\text{min}}$ and $s_{\text{max}}$~\cite{alves2021gamma}. In the case of a space-dependent radiation field, the inverse mean free path along a certain trajectory is easily generalised, in the approximation of $N$ radiation domains: 
\begin{equation}\label{InvMFPtrajectory}
    \lambda^{-1}_\text{traj}(E,z)=\frac{\sum_{i=0}^{N}\lambda^{-1}_{i}(E,z)\cdot l_{i}}{\sum_{i=0}^{N}l_{i}} \,,
\end{equation}
where $\lambda^{-1}_{i}(E, z)$ is the inverse mean free path from Eq.~\ref{InvMFP} computed for the $i$-th photon density, $n_{i}(\epsilon,z)$. Then, $l_{i}$ is the distance covered by the particle in the $i$-th domain. 

The processes ruling EM cascades development in an environment containing background photons are:
\begin{itemize}
    \item pair production \cite{nikishov1961absorption, gould1966opacity}: $\gamma+\gamma_\text{bkg} \to e^{-}+e^{+} $
    \item IC scattering \cite{jones1968calculated}: $e+\gamma_\text{bkg} \to e+\gamma$ 
    \item double pair production \cite{cheng1970cross}: $\gamma+\gamma_\text{bkg} \to  2e^{-}+2e^{+}$ 
    \item triplet pair production \cite{motz1969pair, haug1975bremsstrahlung,mastichiadis1991relativistic, haug2011bremsstrahlung}: $e+\gamma_\text{bkg} \to e+e^{-}+e^{+}$
\end{itemize}
Double and triplet pair production are higher order processes that have been treated in propagation over cosmological distances~\cite{bonometto1972metagalactic,dermer1991effects,mastichiadis1994effect,demidov2009double,ruffini2016cosmic,heiter2018production}. In a generic development of EM cascade, muons or hadrons can be produced for extremely-high~$s$ values in $\gamma + \gamma_{bkg}$ (or $e + \gamma_{bkg}$) interactions~\cite{esmaeili2022ultrahigh}. However, the $s$ variable ranges treated in this work are -- by far -- below the thresholds for muon and charged pion pair productions, viz. $s_{thr} \geq 4m_{\mu / \pi}^{2}c^{4}$, so also for the production of any other heavier hadrons~\cite{whalley2001compilation}. Beside these interactions, charged particles in the cascades suffer from synchrotron energy losses in presence of a magnetic field~\cite{blumenthal1970bremsstrahlung}, which are small in this case. 

\begin{figure}
    \centering
    \includegraphics[width=0.51\textwidth]{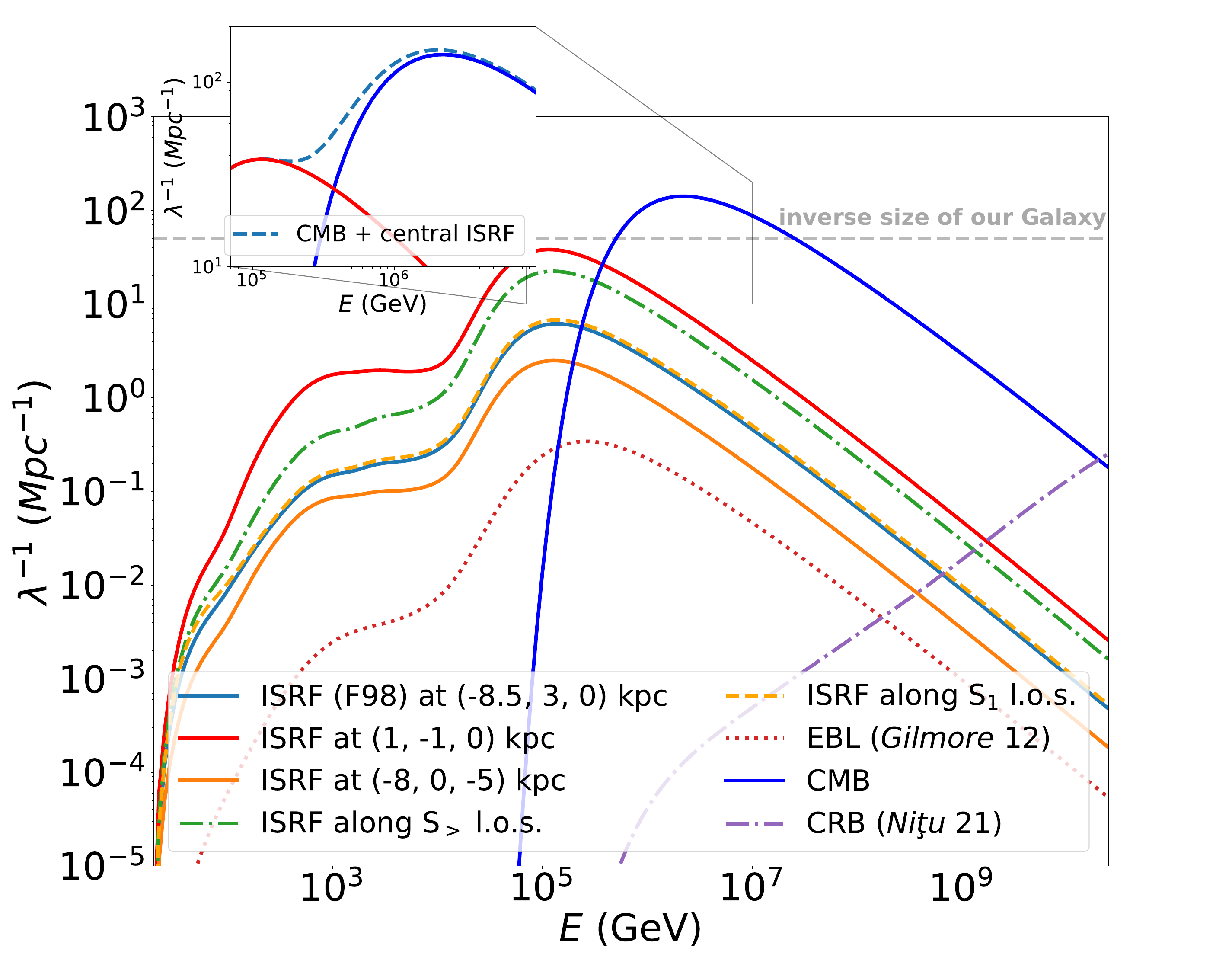}
    \caption{Pair production inverse mean free path from Eq.~\ref{InvMFP} for CRB \textit{(purple dashed-dotted line)} from Ref.~\cite{nictu2021updated}, CMB \textit{(solid blue line)} and EBL \textit{(red dotted line)} from Ref.~\cite{gilmore2012semi}. The ones for the ISRF (F98 model) are computed at three different positions: in the nearby of the galactic center, close to Earth position and $\sim 5 \; \text{kpc}$ out of the galactic plane. The \textit{dashed-dotted green line} is the quantity computed from Eq.~\ref{InvMFPtrajectory} for the line of sight towards the $S_{>}$~source $\sim 25 \; \text{kpc}$ far in Fig.~\ref{fig:GalBfieldSources} (\textit{red} ISRF domains). The \textit{orange dashed line} refers to the line of sight to $S_{1}$~source (\textit{blue} domains). In the inset, the \textit{violet dashed line} is the combined inverse mean free path from CMB and ISRF, the latter in the vicinities of the galactic center.}
    \label{fig:PPrates}
\end{figure}

\begin{figure}
    \centering
    \includegraphics[width=0.51\textwidth]{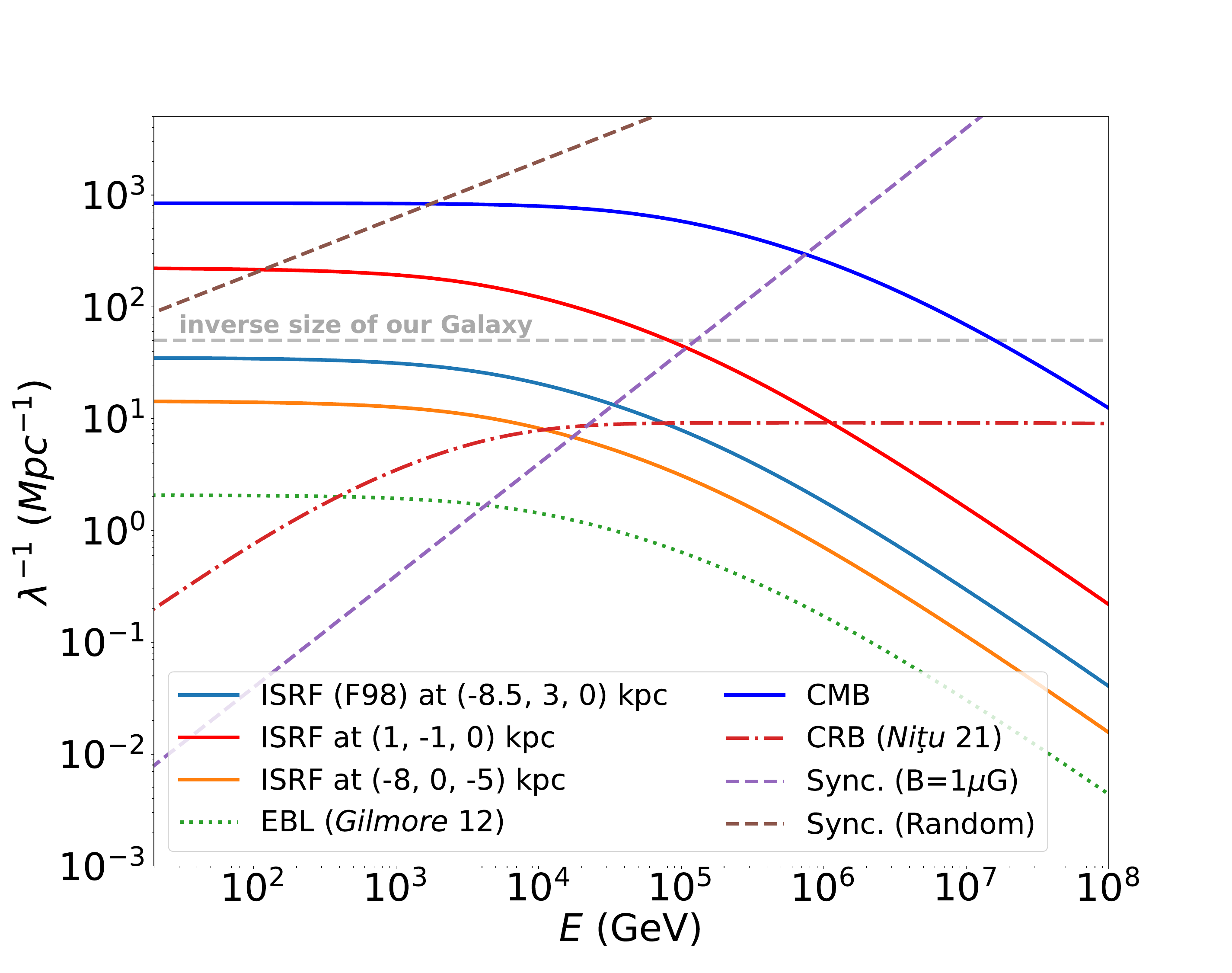}
    \caption{IC inverse mean free path as a function of the electron energy for the treated radiation fields: CRB \cite{nictu2021updated}, CMB, EBL \cite{gilmore2012semi} and three representative ISRF (F98 model). The line called \textit{Sync.} is computed for the synchrotron loss distance travelled in a homogeneous magnetic field. The \textit{Sync. Random} one is the average distance for a randomly oriented magnetic field of 1~$\mu$G: each magnetic field domain has a size of 1 pc~\cite{spurio2018probes}.}
    \label{fig:elEnergyLosses}
\end{figure}

A brief description of the radiation fields permeating our Galaxy follows. 
\paragraph*{\textbf{Cosmic microwave background}} -- The CMB is an isotropic blackbody photon field with a temperature of 2.73~K. It comes from the so called last scattering period in the standard cosmological model~\cite{aghanim2020planck}. Fig.~\ref{fig:PPrates} shows how the CMB pair production inverse mean free path is peaked around $\sim 1 \; \text{PeV}$. Moreover, electrons with energies up to several PeVs mainly up-scatter CMB photons, as the inverse mean free path for IC scattering in Fig.~\ref{fig:elEnergyLosses}. 

\paragraph*{\textbf{Interstellar radiation field}} -- Emission of stars and consequent starlight scattering, absorption and re-emission by the interstellar dust generate the ISRF. It is the galactic counterpart of the EBL. Both ISRF and EBL photon energies range from the infrared to the ultraviolet. Spatial models of the ISRF all over our Galaxy embed together various observations~\cite{gondhalekar1975interstellar, freudenreich1998cobe, robitaille2012self, popescu2017radiation, romero2023predicting}. Two of the most updated 3D models, implemented in the GALPROP code~\cite{Vladimirov_2011}, are reported in Ref.~\cite{porter2017high}. The corresponding benchmark works are Refs.~\cite{freudenreich1998cobe} and~\cite{robitaille2012self}. A further recent work on modelling a 2D axisymmetric ISRF is Ref.~\cite{popescu2017radiation}.

The model from Ref.~\cite{freudenreich1998cobe}, hereafter F98, is employed in this work. Its structure consists in a non-axisymmetric stellar bulge and the stellar and a dust distributions following exponential discs. Its integrated energy density distribution at the galactic plane is shown in Fig.~7~(\textit{right}) of Ref.~\cite{porter2017high}. Contrary to the model of Ref.~\cite{robitaille2012self}, F98 model does not carry galactic arms features, since the average spectral luminosities from galactic arms are taken.

\begin{figure}
    \centering
    \includegraphics[width=0.51\textwidth]{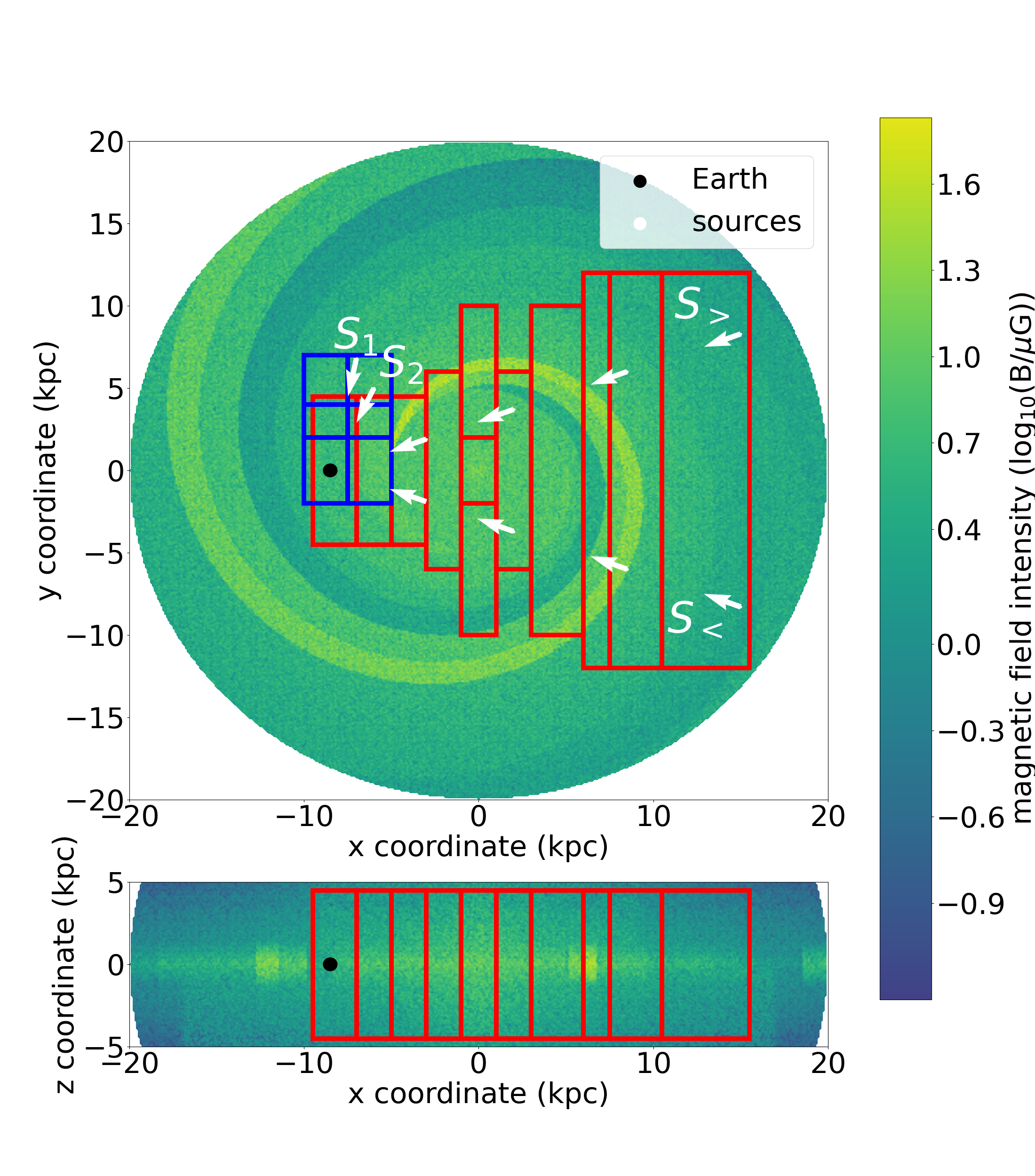}
    \caption{Galactic magnetic field intensity from JF12+~\cite{kleimann2019solenoidal}. The transversal section of the galactic plane (\textit{up}) and from a side (\textit{bottom}) are represented. 
    The observer is represented by the \textit{black circle} at Earth position. Sources are represented as \textit{white arrows} whose heads indicate the emission directions. The ISRF regions implemented in this work are the \textit{red boxes} for $S_{\lessgtr}$~sources and the \textit{blue} ones for $S_{1}$ and $S_{2}$~sources.}  
    \label{fig:GalBfieldSources}
\end{figure}

In the current CRPropa version, position-dependent radiation fields are not implemented. For this reason, to approximate the ISRF model, lines of sight to the sources are divided into different ISRF density regions, i.e. the red/blue boxes in the simulation setup shown in Fig.~\ref{fig:GalBfieldSources}, employing the \texttt{RestrictToRegion} module in CRPropa. In each domain, the ISRF reference density is taken as the closest one to the lines of sight. Eq.~\ref{InvMFPtrajectory} is adopted to compute the total inverse mean free path for pair production along the line of sight in Fig.~\ref{fig:PPrates}.

\paragraph*{\textbf{Cosmic radio background}} -- 
It is the background photon field due to the integrated history of galaxy and star emissions. These lower energy photons usually dominate the attenuation of gamma rays with energies larger than $10^{18}$~eV. The model adopted in this work is the one reported in Ref.~\cite{nictu2021updated}, a refinement of a previous study~\cite{protheroe1996new}. The novelty of Ref.~\cite{nictu2021updated} lies in the more accurate treatment of the radio processes occurring in star-forming and radio galaxies, as well as their morphology and evolution. As shown in Fig.~\ref{fig:elEnergyLosses}, IC off CRB photons starts to be relevant for hundreds-of-TeV electrons, dominating at energies beyond 100~PeV. 
\subsection{Galactic magnetic field model}\label{sub_magnModl}

Throughout the development of EM cascades, astrophysical magnetic fields play a role in deflecting electrons and causing them to lose energy through synchrotron emission. 

In reproducing galactic environments, the GMF model employed in this paper is the one reported in Ref.~\cite{kleimann2019solenoidal} (JF12+ hereafter), an implemented version of Ref.~\cite{jansson2012new} -- the former only improves the out-of-plane GMF and ensures the magnetic flux conservation. It comprises three components: the large-scale regular field, small-scale turbulent fields, with coherence length of 100~pc or smaller, and the striated random field, aligned over larger scales, but with varying strength and sign. The colour maps of Fig.~\ref{fig:GalBfieldSources} represent the magnetic field intensity at the galactic plane for the three contributions combined. The regular component is, in turn, made up of three items: the disc, the toroidal halo, and the out-of-plane component. 

The interplay between regular and turbulent structures might be relevant in interpreting the morphology of the arrival direction of the ``deflected" gamma rays. Indeed, observed gamma-ray count maps are expected to be shaped according to the specific magnetic field configuration traversed by electrons in EM cascades. In case of no misalignment between the observer line of sight and the emission direction of the source, if $\langle B_\text{turb}/B_\text{reg}\rangle_\text{los} \gg 1$, the count map is expected to show a homogeneous halo around the core of the emission. This fact is due to the procedure of stacking many magnetic field realisations, averaging out random component contributions. In the opposite case or, at least, if $\langle B_\text{turb}/B_\text{reg}\rangle_{\text{los}} \sim 1$, the observed counts map is expected to show asymmetric features due to the effects of the regular component on the charged particles in the EM cascade, in a similar fashion to what happens for extragalactic sources~\cite{alvesbatista2017d}. In App.~\ref{AppBfield} magnetic field structures and components are described in more details.

\subsection{Spherical observer \& gamma-ray sources}

The observer, \textit{black circle} in Fig.~\ref{fig:GalBfieldSources}, is defined as a sphere whose origin lies at the nominal Earth position, in agreement to JF12+. In our simulations, the observer radius is taken as the geometrical one, i.e. $R_\text{obs}=D\cdot \tan(\phi+\delta)$, with $\phi\sim 1^{\circ}$ is the half-aperture of the emission cone, $\delta \ll \phi$ and $D$ the distance source-observer. An in-depth discussion about the observer sphere size is in App.~\ref{AppObsSize}.   

The ten simulated sources are placed around the Galaxy as the \textit{white arrows} in Fig.~\ref{fig:GalBfieldSources}, all of them pointing towards Earth. There are two sets of sources mirror-symmetric with respect to the x-axis, hereafter $S_{\lessgtr}$ because of their $\lessgtr 0$ y-coordinate. Each contains four sources, with a line of sight that goes through the inner regions of our Galaxy. For these two sets, the ISRF domains implemented are delimited by \textit{red lines} in Fig.~\ref{fig:GalBfieldSources}. The other two sources are 7~kpc (alias $S_{1}$) and 5.6~kpc ($S_{2}$) far from Earth: the related ISRF regions are bounded by \textit{blue lines} in Fig.~\ref{fig:GalBfieldSources}. The lines of sight cross, respectively, one and two galactic spiral arms, seen in the GMF of Fig.~\ref{fig:GalBfieldSources}. Interestingly, the distance and line of sight -- far from the galactic center -- of the $S_{1}$ source resemble the location of 1LHAASO J2002+3244u, one of the multi-TeV gamma-ray sources in the first LHAASO catalog~\cite{cao2024first}.

In Fig.~\ref{fig:PPrates}, the weighted inverse mean free paths from Eq.~\ref{InvMFPtrajectory} are computed for the lines of sight related to the farthest source among $S_{>}$~set, $\sim 25 \; \text{kpc}$ distant, and the source $S_{1}$. The former, in particular, is few times smaller than the inverse radius of our Galaxy at around 100~TeV. 

Some properties we look for in choosing the position of sources are:
\begin{itemize}[noitemsep,nolistsep]
    \item the distance source-observer, to be compared with the length scales of the interactions;
    \item the turbulent \textit{vs} regular component ratio along the line of sight, thus the GMF spiral arms crossed by electrons in EM cascade; 
    \item the ISRF energy densities in the regions traversed by the photons.  
\end{itemize}

\section{Results and discussion}\label{result_sec}

This section reports the main results from the simulation of gamma rays injected by each of the ten sources defined above. It is divided into two parts. Sec.~\ref{Spectra} deals with the energy spectra, whereas Sec.~\ref{haloes} concerns gamma rays arrival direction. 
\subsection{Energy spectra}\label{Spectra}
\begin{figure*}
    \begin{minipage}{0.49\textwidth}
    \centering
    \includegraphics[width=\linewidth]{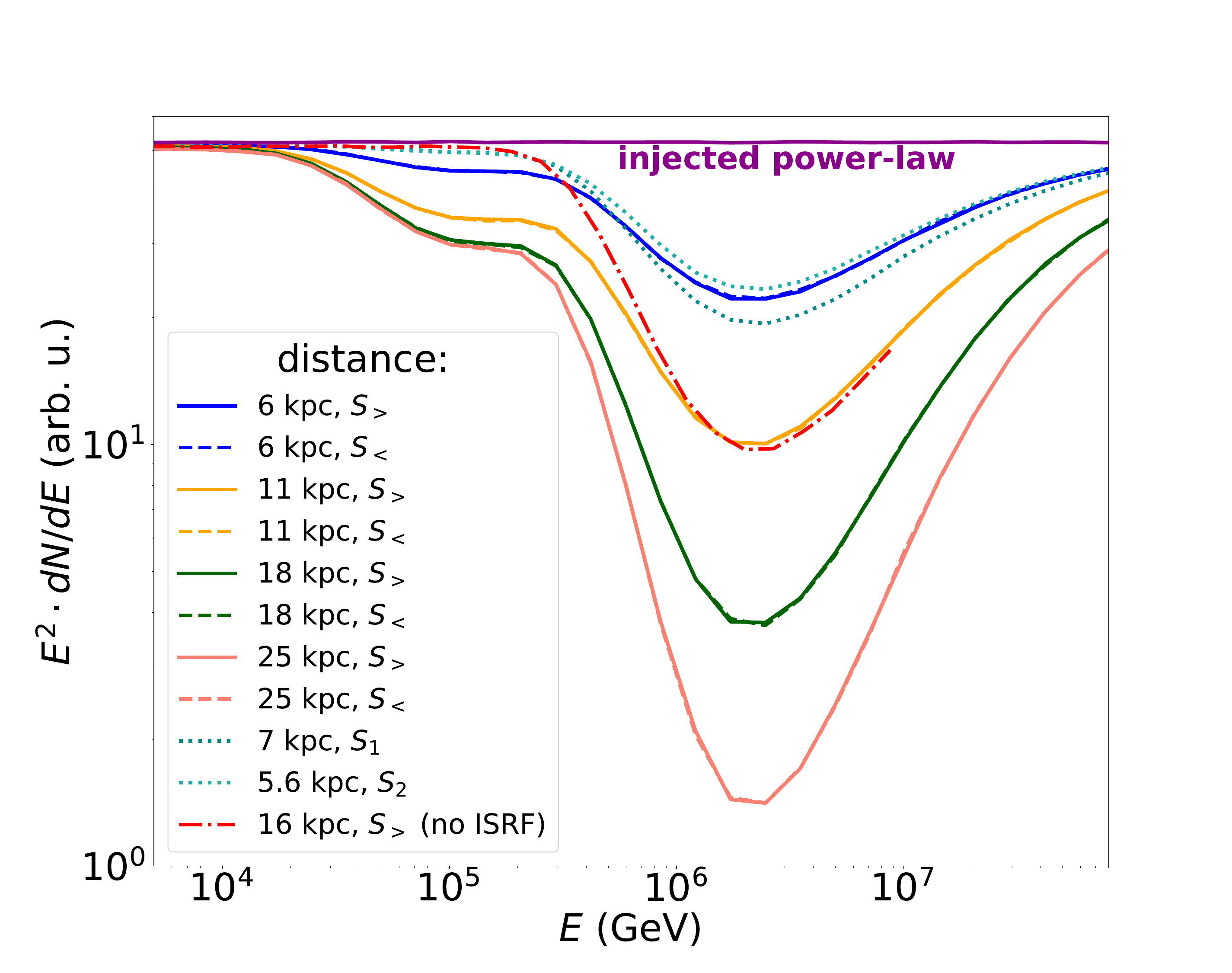}

    \captionsetup{labelformat=empty}
    \caption*{(a) Power-law with a spectral index 2.}
    \end{minipage}
    \hfill
    \begin{minipage}{0.49\textwidth}
    \centering
    \includegraphics[width=\linewidth]{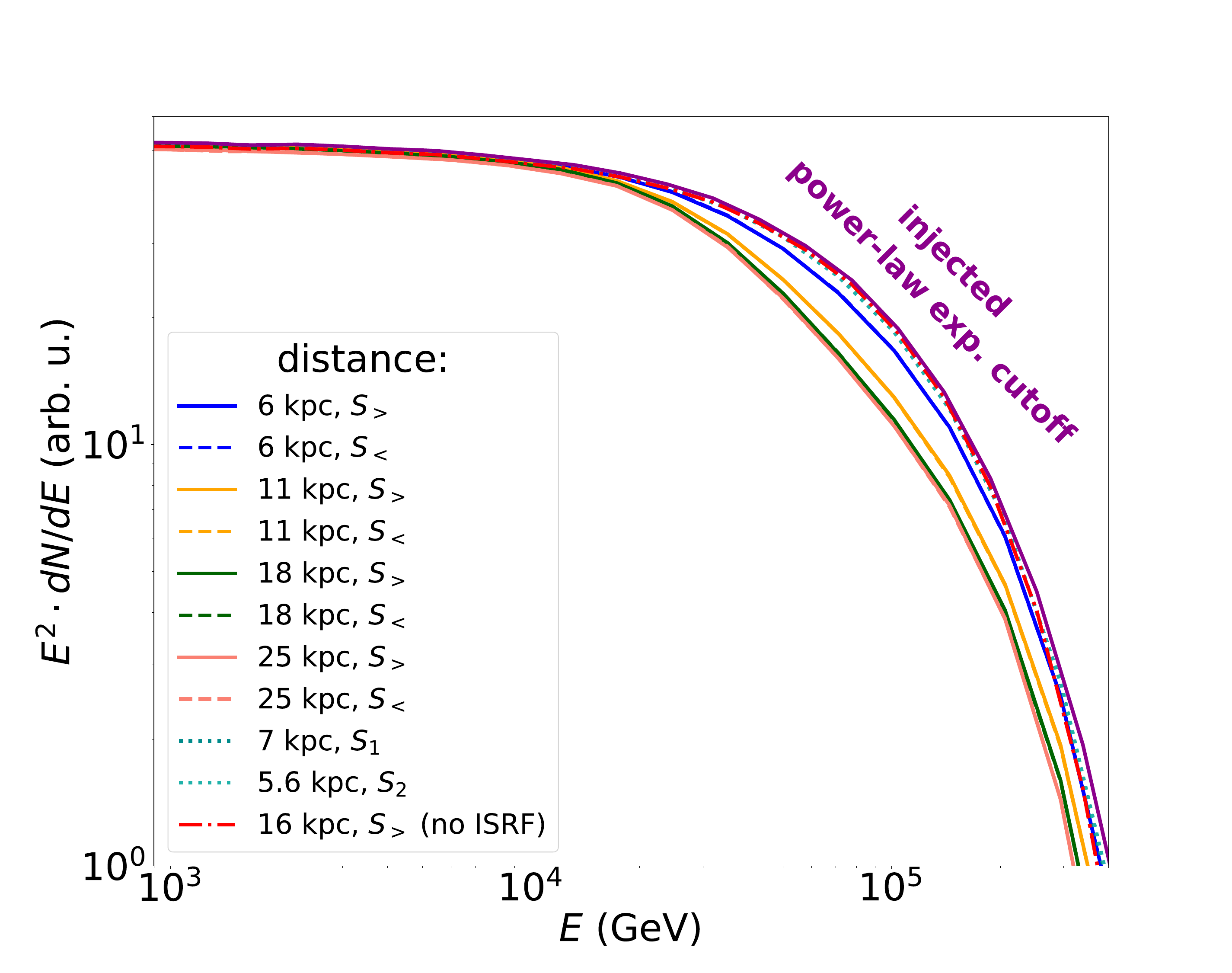}
    
    \captionsetup{labelformat=empty}
    \caption*{(b) Power-law ($\alpha$=2) with an exponential cut-off at 100 TeV.}
    \label{fig:energySpectracutOff}
    \end{minipage}
    \caption{Energy spectra for the 10 sources of Fig.~\ref{fig:GalBfieldSources}, both for an injected energy spectrum described by a power law, Eq.~\ref{power-lawEq}, in~(a) and by an exponential cut off, Eq.~\ref{power-lawCOEq}, in~(b). The spectral energy distributions for the two sets, $S_{>}$ (\textit{continuous lines}) and $S_{<}$ (\textit{dashed}), overlap at a fixed distance. The observed energy spectra from $S_{1}$ and $S_{2}$~sources are represented by \textit{dotted lines}, respectively \textit{cyan} and \textit{sea green}. As a reference, the injected energy spectra are given by the \textit{continuous dark magenta lines}. The energy spectrum for a simulation with the EBL replacing the ISRF is also shown (\textit{dashed-dotted red line}). In this latter scenario, we simulate gamma rays \textit{only} up to 10~PeV.} 
    \label{fig:energySpectra}
\end{figure*}

The ten sources energy spectra are in Fig.~\ref{fig:energySpectra}(a) for a simple power-law intrinsic spectrum profile. The prompt emission is modelled as:
\begin{equation}\label{power-lawEq}
   \diff{N}{E}\propto E^{-\alpha} \,,
\end{equation}
with spectral index $\alpha=2$. In Fig.~\ref{fig:energySpectra}(a) the observed spectra with ISRF decrease at $\sim 10 \; \text{TeV}$ by a factor $\sim 2$, at most, for $S_{\lessgtr}$~sources. Then, after a plateau, they start going down again. The absolute minima in the energy spectra, or equivalently the maximum absorptions, are between 1 and 2~PeV. They are entirely caused by pair production. Double pair production does not contribute to gamma-ray absorption over galactic distances.

The two main spectral features are better understood by looking at Fig.~\ref{fig:PPrates}: the ISRF inverse mean free path for pair production is characterised by a two-peaks profile. The local maximum inverse mean free path at $\sim 1 \; \text{TeV}$ is compatible with the first absorption in the energy spectra at $\sim 10 \; \text{TeV}$. The absolute maximum in the ISRF inverse mean free path at $\sim 100 \; \text{TeV}$ has to be combined with the CMB one, placed at $\sim 10 \; \text{PeV}$. The maximum absorption is located at $\sim 2 \; \text{PeV}$.

The lines of sight of $S_{\lessgtr}$~sources go through the vicinities of the galactic center, where the highest ISRF energy densities are observed. Even though the energy spectra profile is similar changing the source distance, the amount of absorption is distance-dependent. Furthermore, at each fixed distance, energy spectra of the $S_{>}$~sources cannot be distinguished from their specular ones, i.e., $S_{<}$~sources. The two sources along galactic spiral arms, $S_{1}$ and $S_{2}$, present a tangible absorption only above 200~TeV, \textit{dotted lines} in Fig.~\ref{fig:energySpectra}(a), until reaching the minimum around 2~PeV. Likewise in this case, the amount of absorption depends on the distance.

In the case without ISRF, thus including the EBL from Ref.~\cite{gilmore2012semi}, the absorption starts only at 1 PeV due to the interaction of gamma rays with CMB photons (\textit{dashed-dotted red line} in Fig.~\ref{fig:energySpectra}). The minimum in the spectrum, or maximum absorption, is at $\sim 2 \; \text{PeV}$ as well. 
Simulating sources that inject gamma rays from 100~GeV to 100~PeV is an instructive situation in order to characterise gamma-ray absorption due to propagation effects in our Galaxy. Nevertheless, no astrophysical sources are expected to emit according to a pure power-law spectrum over such a wide energy range. Recent gamma-ray observations till few PeVs are usually modelled introducing a cut-off function, whether sharp, log-parabola, or an exponential cut-off, as e.g. in Ref.~\cite{cao2021ultrahigh}.

The differential flux dependence for an exponential cut-off spectral power-law model is:
\begin{equation}\label{power-lawCOEq}
\diff{N}{E} \propto E^{-\alpha} \exp\left(-\dfrac{E}{E_0}\right) \,,
\end{equation} 
where $E_{0}$ is the cut-off energy. Note that a simple power-law, thus Eq.~\ref{power-lawEq}, is recovered for $E_{0}\to\infty$. Eq.~\ref{power-lawCOEq} is merely a phenomenological formula that may vary depending on the different types of sources. In Fig.~\ref{fig:energySpectra}(b) the energy spectra with $E_{0}=100$~TeV are shown. With the introduction of an exponential cut-off, the difference between the case with and without ISRF is remarkable for the farthest $S_{\lessgtr}$ source, i.e. for distances $\gtrsim 11 \; \text{kpc}$. For the two $S_{\lessgtr}$ sources $6 \; \text{kpc}$ distant is softly below the injected spectrum, while the spectra of the $S_{1}$ and $S_{2}$ sources are not affected by propagation effects. If ISRF is not at play, no attenuation is at work for this injection model.  

\begin{figure}
    \centering
    \includegraphics[width=0.51\textwidth]{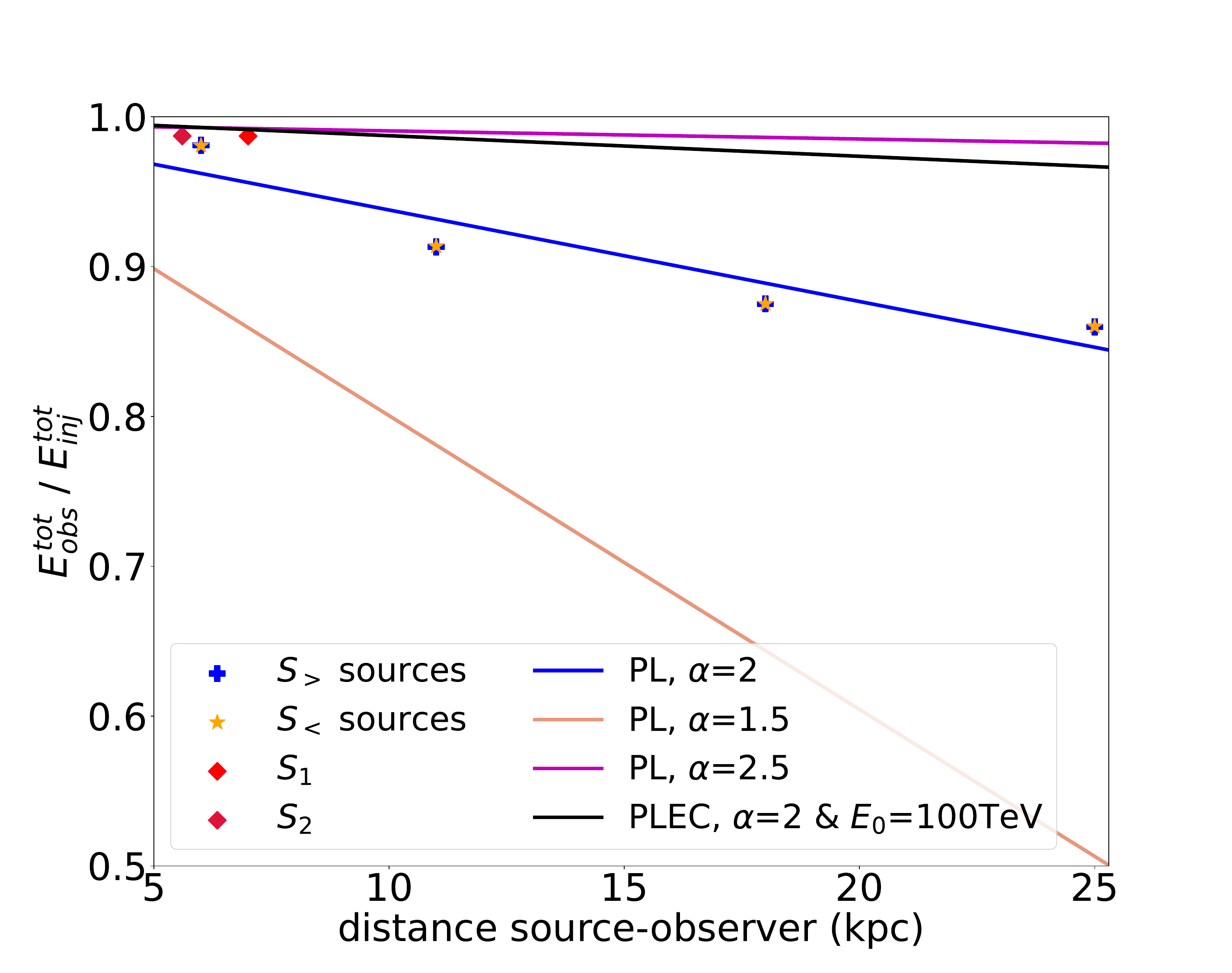}
    \caption{Ratios between the total observed energy and the
injected one. Data points are for the 10 sources considered in this work, and for the case in which the injection model is a power law with $\alpha=2$. The \textit{blue line} corresponds to the fit to the $S_{>}$~set; see Table~\ref{table:absorptionRates}. Other fits are also shown in different color lines for the different injection models of the corresponding $S_{>}$~data points, as given in the legend.}
    \label{fig:energyPlot}
\end{figure}

The ratios between the total observed energy and the injected one in dependence to the source-observer distance are shown in Fig.~\ref{fig:energyPlot}, for different emission models. Data points are computed for the case of the power law with spectral index 2. The \textit{blue line} is the fit referring to $S_{>}$~sources (\textit{blue points}). A similar result is obtained for $S_{<}$~sources (\textit{orange stars}), consistently with the energy spectra in Fig.~\ref{fig:energySpectra}. This reference power-law model is compared with larger and smaller spectral indexes and the case with a cut off at 100~TeV, as in Fig.~\ref{fig:energySpectra}(b). Table~\ref{table:absorptionRates} provides the absorption ratio lengths of the four emission models. The softer the injection model, the smaller the amount of absorbed and/or dispersed energy is. This trend is due to the smaller amount of higher energy photons for larger spectral indexes, i.e. the most affected by absorption and ``deflection".
\begin{table}
  \centering
  \caption{In the last column, the slopes of the linear fits in Fig.~\ref{fig:energyPlot}. It reflects the dependence of the distance on the ratio between observed and injected energy. Each row corresponds to a different injection model -- PL stands for \textit{power-law}, in Eq.~\ref{power-lawEq}, while PLEC for \textit{power-law exponentially cut off}, Eq.~\ref{power-lawCOEq}.}
\begin{tabular}{cccc}
\hline
 spectrum type & $\alpha$ & $E_0 \; \text{[TeV]}$ & absorption [$10^{-3} / \text{kpc}$] \\
\hline
 PL & 1.5 & --- & 19.6 \\
 PL & 2.0 & --- & 6.1 \\
 PL & 2.5 & --- & 0.5 \\
 PLEC & 2.0 & 100 & 1.4 \\
\end{tabular}
\label{table:absorptionRates}
\end{table}
\subsection{Arrival directions}\label{haloes}

Galactic gamma-ray ``deflection'' is studied in terms of the difference between arrival and initial momenta for each observed event. All spatial count maps are shown in App.~\ref{AppCounntMaps}. Despite a virtually negligible contribution from triplet pair production in the interactions between EM cascade electrons and background photons over galactic distances, IC scattering is the dominant process. Indeed, IC is responsible for the production of secondary gamma rays.

Spatially extended ``haloes'' start forming at a distance of $\sim 11 \; \text{kpc}$, considering the sources in the $S_{\lessgtr}$~sets. Indeed the count maps in Fig.~\ref{fig:map_plots_close}(a)--(b), i.e. the ones for the closest sources in the $S_{\lessgtr}$ set, only $\sim 6 \; \text{kpc}$ far away, present only a few sparse counts around the centered, point-like sources. Moving $\sim 5 \; \text{kpc}$ farther, i.e. Fig.~\ref{fig:map_plots_close}(c)--(d), defined halo shapes are already noticed. The two halo shapes are different for the considered $S_{\lessgtr}$ cases at a fixed distance. Panels (e)--(f) in Fig.~\ref{fig:map_plots_far} show the count maps for the two $S_{\lessgtr}$~sources $\sim 18 \; \text{kpc}$ distant, whereas panels (g)--(h) in the same Fig.~\ref{fig:map_plots_far} are the ones corresponding to the sources $\sim 25 \; \text{kpc}$ far. Overall, the larger the distance, the more spread counts are in the selected angular window. The trend is due to the energy threshold of the simulations combined with the dispersion effects of the magnetic field on electrons. On the one side, cascade particles with energies lower than 100~GeV are no more tracked. On the other one, cascades of the farthest $S_{\lessgtr}$ sources go through highly magnetised regions, causing stronger electron deflections. In this case the magnetic field might cause a sort of \textit{screening effect} for the observer at Earth.

Spatial count maps for the other two sources, $S_{1}$ and $S_{2}$, are respectively shown in Fig.~\ref{fig:PP0_7kpc} and Fig.~\ref{fig:PP0_5.6kpc}. The two shapes are strongly irregular both in orientation and in counts distribution around the point-like source at the center. As pointed out in Sec.~\ref{sub_magnModl}, these effects are attributed to the peculiar structure of the magnetic field along the line of sight. In particular, we recall that Fig.~\ref{fig:Bfiel_along_los_manysp} and~\ref{fig:Bfiel_along_los_spiral} in App.~\ref{AppBfield} show, respectively, the magnetic field structures along the line of sight of the $S_{>}$~source $\sim 6 \; \text{kpc}$ far from Earth, and the one of the $S_{1}$~source. If the latter line of sight crosses only one regular magnetic field region, the former one is greatly dominated by the turbulent component. In fact, the mean ratio along the line of sight, i.e. $\langle B_\text{turb}/B_\text{reg}\rangle_\text{los}$, related to the $S_{1}$~source is about 2.3. In the scenario for the $S_{>}$~source, $\sim 6 \; \text{kpc}$ distant, the ratio is $\sim9.5$. For all the $S_{\lessgtr}$~sources, the line of sight magnetic field structure is extremely \textit{turbulent-dominated} as well.  

\begin{figure*}
     \begin{minipage}[t]{0.49\textwidth}
         \centering
         \includegraphics[width=\linewidth]{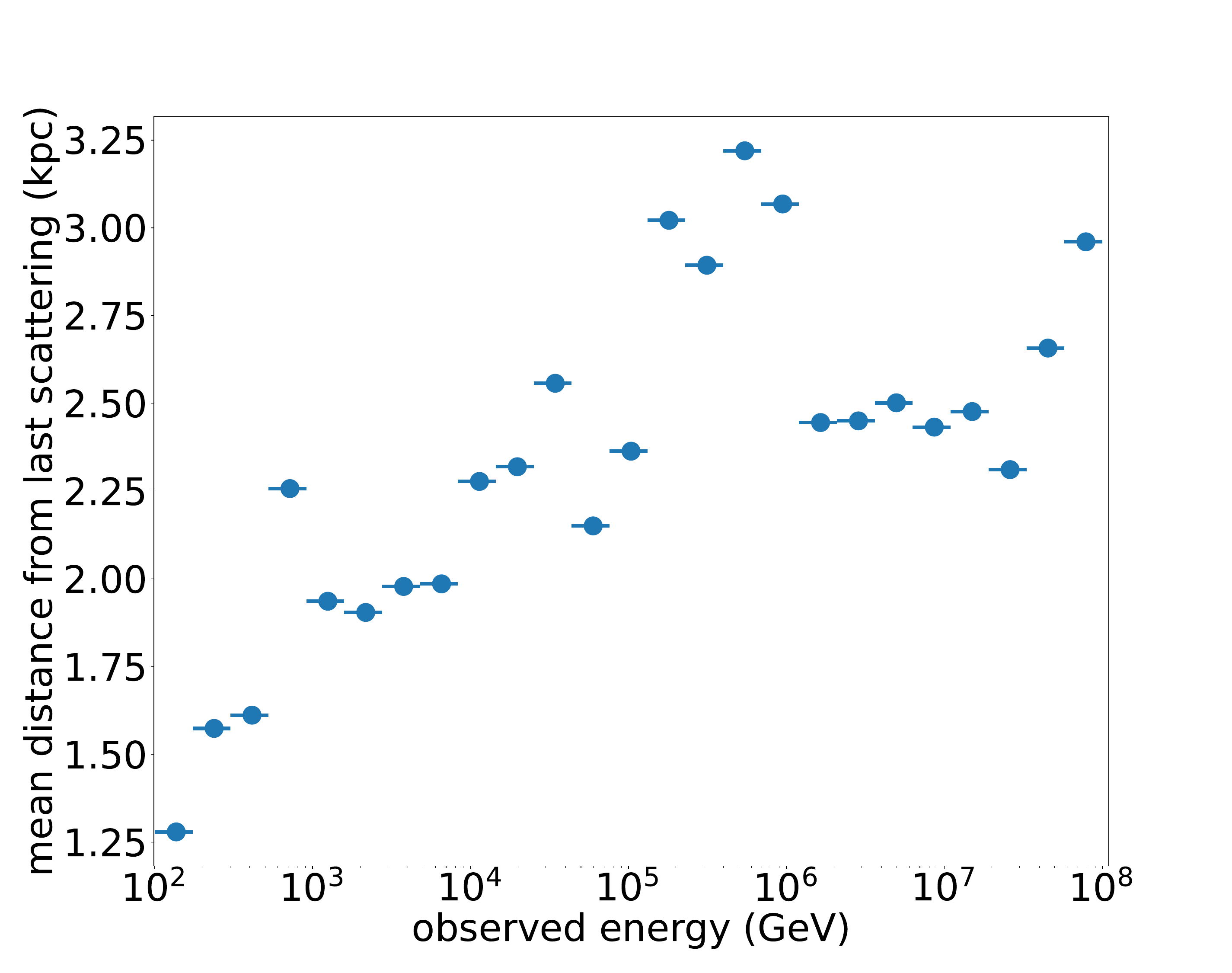}

         \captionsetup{labelformat=empty}
         \caption*{(a) $S_{>}$~source; $D\sim 11 \; \text{kpc}$.}
    \end{minipage}
    \medskip
    \begin{minipage}[t]{0.49\textwidth}
         
         \includegraphics[width=\linewidth]{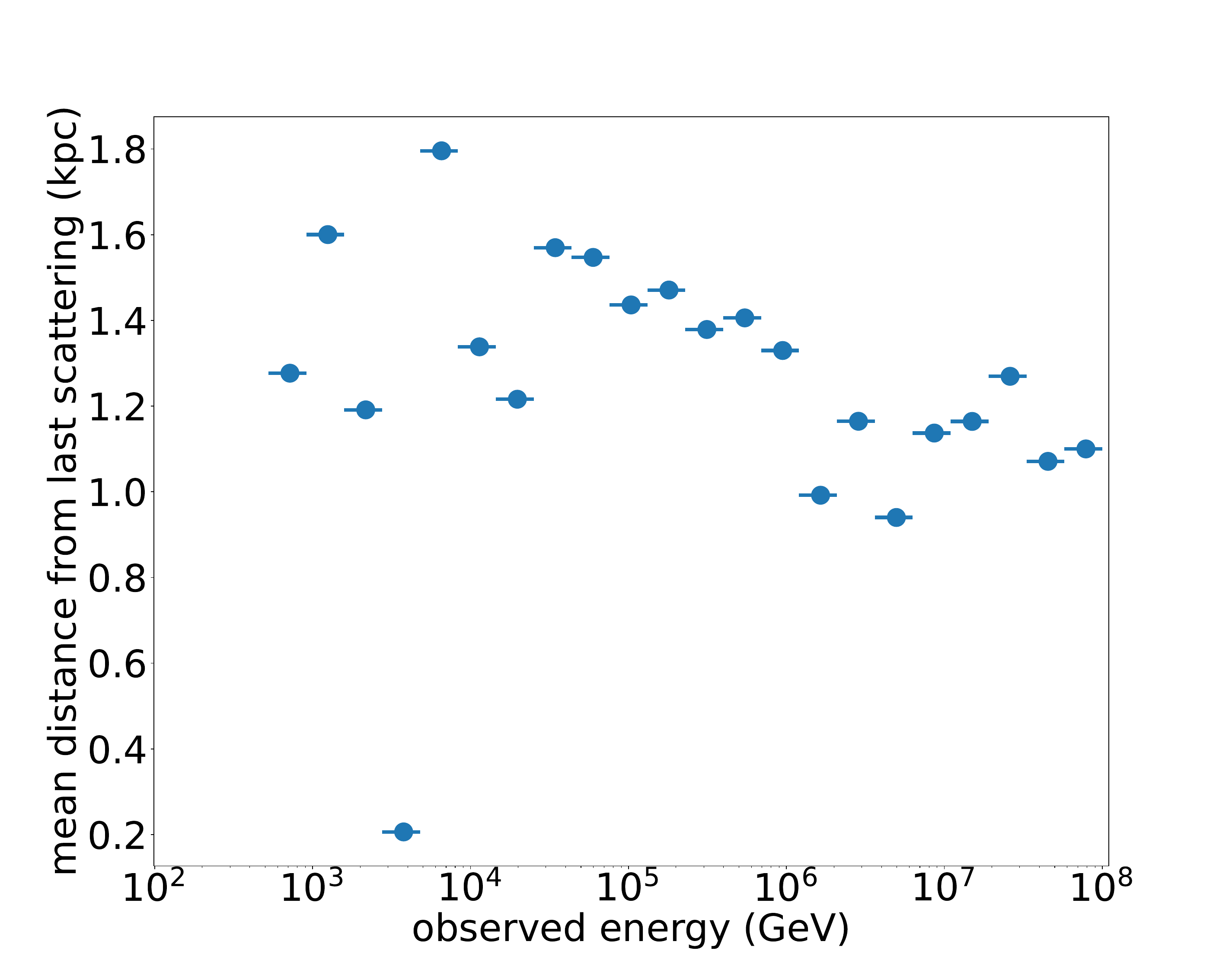}
         \captionsetup{labelformat=empty}
         \caption*{(b) $S_{1}$~source; $D\sim 7 \; \text{kpc}$. To notice that no secondary gamma rays are observed below 600~GeV.}
    \end{minipage}
     \caption{\textit{Last scattering} plots of IC-produced gamma rays.}
     \label{fig:ls_plots}
\end{figure*}

A useful tool to better understand the halo morphologies comes from the so-called ``\textit{last scattering}'' plots. These show the distance between the points at which secondary gamma rays are produced and observed. \textit{Last scattering} plots are reported in Fig.~\ref{fig:ls_plots} for the $S_{>}$~set, $\sim 11 \; \text{kpc}$ far, and for the $S_{1}$~source. Substantial changes among them are observed, which are related to the magnetic field structures previously described, as well as the different ISRF energy densities.

The $S_{1}$ highly-shaped count map of Fig.~\ref{fig:PP0_7kpc} is better understood by combining the corresponding magnetic field configuration of Fig.~\ref{fig:Bfiel_along_los_spiral} in App.~\ref{AppBfield} and the \textit{last scattering} plot in Fig.~\ref{fig:ls_plots}(b). Secondary observed photons are produced at distances less than $\sim 1.8 \; \text{kpc}$ from the observer. This means that electrons, before up-scattering photons, travel distances of few parsecs in a magnetic field in which the regular magnetic component is comparable to the turbulent one. We note that the turbulent component is more prominent only in the nearby of $S_{1}$~source, taking into account the line of sight to the observer. On the other hand, the count map of the $S_{>}$ source in Fig.~\ref{fig:map_plots_close}(c) exhibits a more uniform halo around: this is due to magnetic field structures akin to the one rendered in Fig.~\ref{fig:Bfiel_along_los_manysp} in App.~\ref{AppBfield}. The corresponding \textit{last scattering} plot of Fig.~\ref{fig:ls_plots}(a) presents a different profile, reliant on both line of sight and distance to the observer. 

\begin{figure*}
    \begin{minipage}[t]{0.49\textwidth}
    \centering
    \includegraphics[width=\linewidth]{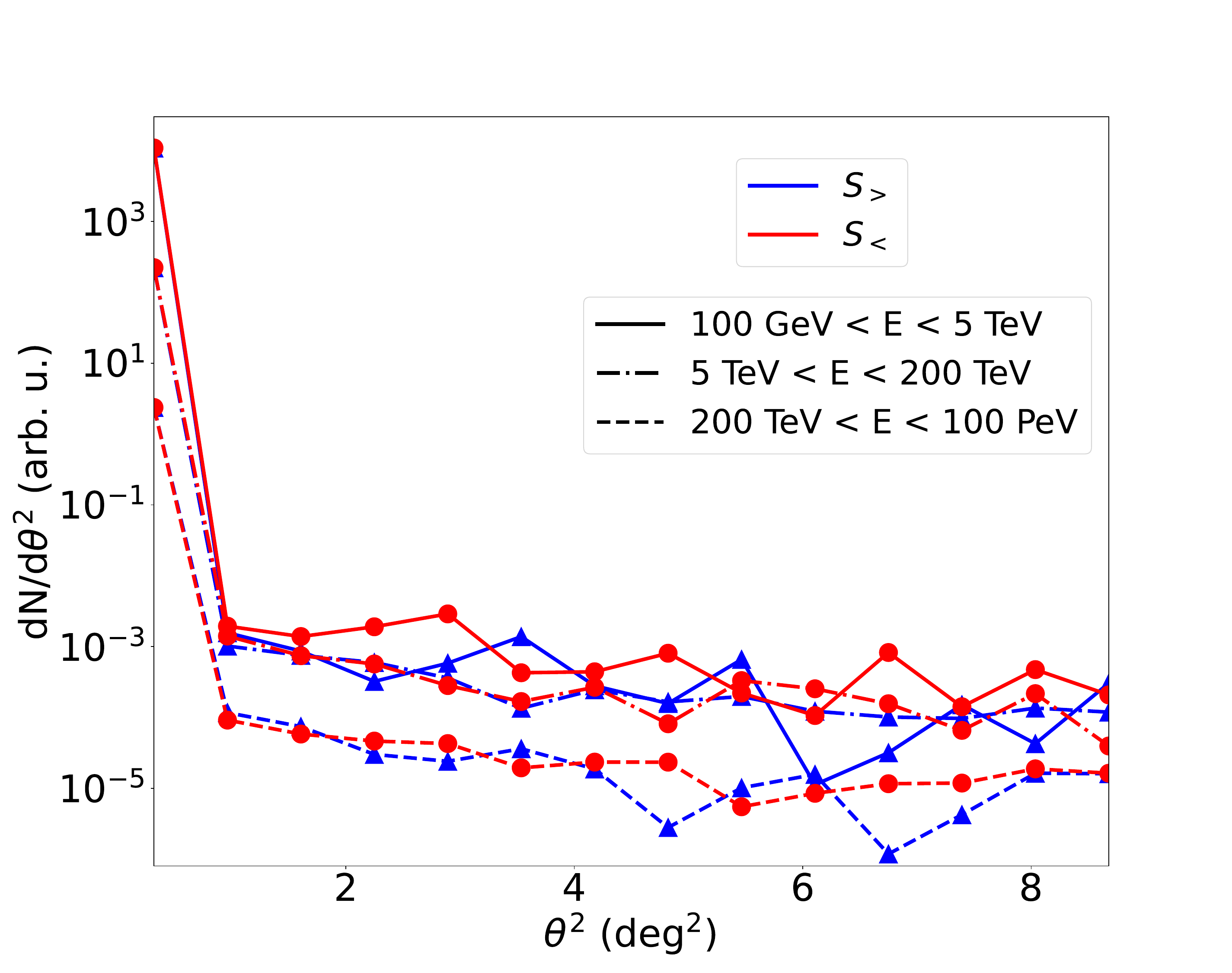}
    \captionsetup{labelformat=empty}
    \caption*{(a) Set of $S_{\lessgtr}$~sources, $D\sim11~\text{kpc}$.}
    \end{minipage}
    \medskip
    \begin{minipage}[t]{0.49\textwidth}
    \centering
    \includegraphics[width=\linewidth]{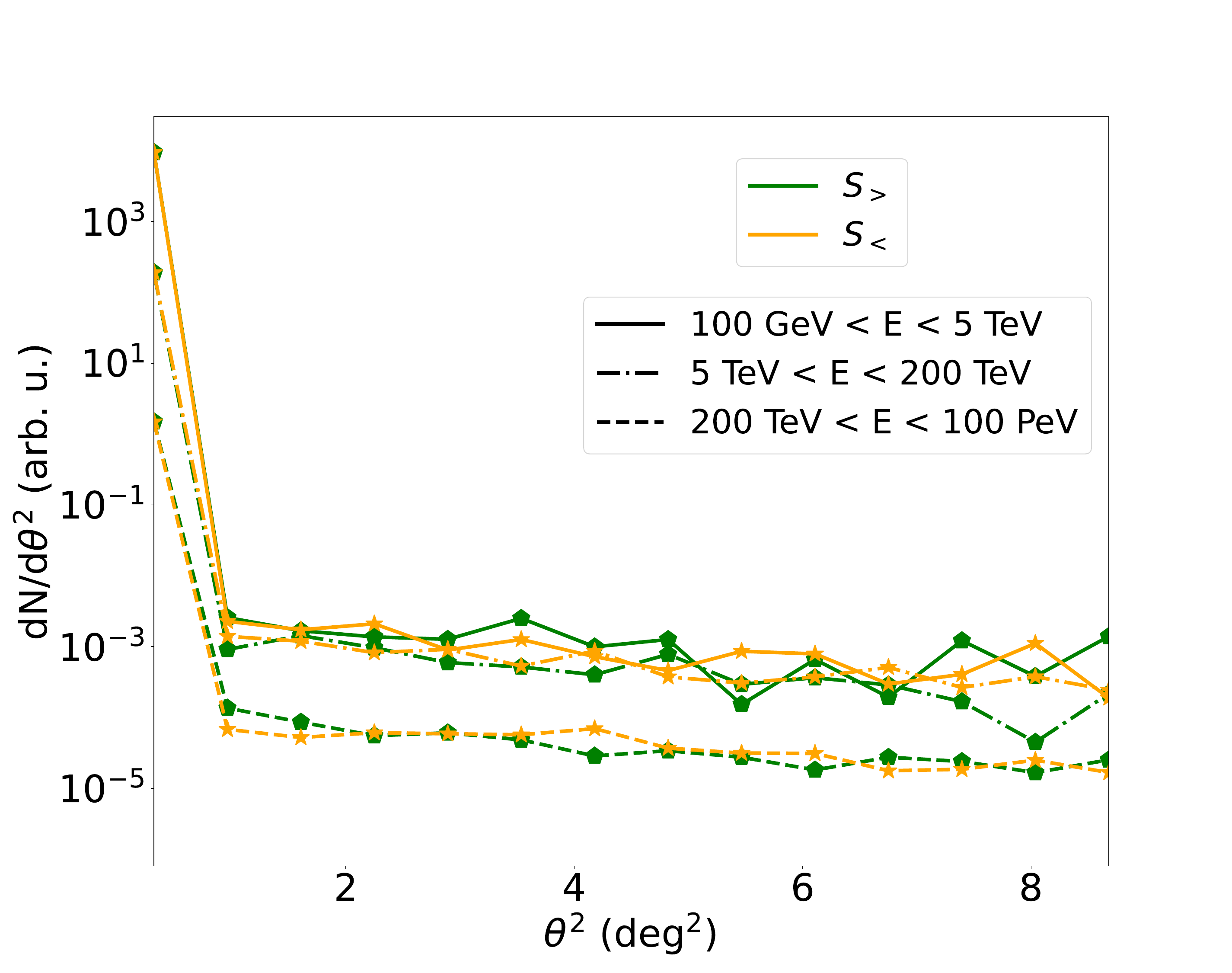}
    \captionsetup{labelformat=empty}
    \caption*{(b) Set of $S_{\lessgtr}$~sources, $D\sim18~\text{kpc}$.}
    \end{minipage}
    \caption{\textit{Surface brightness} profile depicted in three different energy ranges as shown in the legend, for two sources of the $S_{\lessgtr}$~sets located at 11~(a) and 18~(b)~kpc from Earth. Colours refer to sources, while linestyles do so for energy ranges. The reference injection model is a power law with $\alpha=2$.}
    \label{fig:SB_endep}
\end{figure*}

Once we have characterised gamma-ray arrival morphologies due to propagation effects alone, it becomes crucial to determine haloes energy dependence as well. The \textit{surface brightness} profiles for $S_{\lessgtr}$~sources at distances of $\sim 11$ and $\sim 18 \; \text{kpc}$ from the observer are shown, respectively, in Figs.~\ref{fig:SB_endep}(a)--(b). The deflection is defined as $\theta =\arccos\big(\vec{P}\cdot\vec{P_{0}}\big)$, with $\vec{P}$ and $\vec{P_{0}}$ direction at the observer and at emission. In the four cases reported, counts around the source are distributed forming a ``plateau'' at every considered energy range. Yet, its contrast, i.e. the ratio between the centered point-like source over haloes counts, is, at least, of five orders of magnitude. Halo counts are mainly constituted by gamma rays in the energy range between 5~and 200~TeV, given the considered power-law model with $\alpha=2$. No substantial differences are spotted between the specular sources at fixed distances.


\section{Conclusion and prospects}\label{conclPersp}

The aim of this work is to investigate the role of propagation effects on gamma rays traveling through our Galaxy, determined by the environmental characteristics, i.e. background photons and magnetic fields. These drive the development of EM cascades, triggered by gamma rays with energies larger than tens of TeV. 

Depending on the specific source position in the Galaxy, gamma-ray fluxes detected at Earth are found to be absorbed and ``deflected". The energy spectra shown in Fig.~\ref{fig:energySpectra} are evidence of absorptions caused by the combination of two photon backgrounds: the CMB and the ISRF. It starts to be relevant at $\sim 10 \; \text{TeV}$. In the absence of the ISRF, energy spectra show an absorption at $\sim 1 \; \text{PeV}$ due only to the CMB. The power-law energy spectrum profiles are in a qualitative agreement with the \textit{survival probabilities} computed for gamma-ray absorption in Refs.~\cite{moskalenko2006attenuation,zhang2006very,vernetto2016absorption,popescu2017radiation,porter2018galactic}, in which ISRF and CMB photons are taken into account. The novelty of our simulations of galactic gamma-ray sources lies in the spectral and spatial investigation of EM cascades initiated by such highly energetic fluxes, through MC simulation tools. 

Taking into account the ISRF, the amount of absorbed and/or ``deflected'' gamma rays changes in dependence to the distance source-observer considering sources along the same line of sight. By assessing the ratio between the total observed and injected energy, we quantify a direct consequence of the ISRF impact on actual observables. Out of it, an absorption inverse length is inferred for different source emission models, in Fig.~\ref{fig:energyPlot} and Tab.~\ref{table:absorptionRates}. This absorption is found to be directly related to the highest energy photons weights in the spectra, i.e. the larger the spectral index (i.e. the softer the spectrum), the smaller is the absorption per length.

The discussion on the arrival direction maps in Sec.~\ref{haloes} suggests the complicated detectability of spatially extended haloes around point-like sources due to propagation effects alone -- even though they do depend on the intrinsic gamma-ray spectrum of the source. The count maps showed in this work reveal the high contrast -- at least four or five orders of magnitude -- between the centered point-like sources and the deflected counts around. Yet, two interesting considerations are drawn. The first one concerns the role of the GMF and how it affects the arrival directions of gamma rays. In the case of galactic TeV-extended sources, e.g. \cite{sudoh2019tev, lopez2022gamma}, propagation effects might be relevant in enhancing the extension, slightly deforming the shape itself. The latter feature is due to the complex structure of the GMF between source and observer. How much these effects impact current and future gamma-ray observations is left for a future work. The second consideration is about the influence of ``deflected'' gamma rays in our observation. We enquire whether the angular spreading of gamma rays due to propagation effects contribute to the galactic diffuse gamma-ray background~\cite{bloemen1989diffuse, fornasa2015nature, neronov2020galactic, cao2023measurement, fang2023decomposing, yan2024insights, alfaro2024galactic}, considering the so far observed multi-TeV sources of our Galaxy, viz. the ones in LHAASO and HAWC catalogs~\cite{albert20203hwc,cao2024first,albert2024observationgalacticcenterpevatron}. In a broader multi-messenger view, a potential impact on the galactic diffuse gamma-ray background might improve the diffuse neutrino background expectations from gamma rays~\cite{fang2024neutrinos}. Another interesting concern regards the role of the electrons generated in the EM cascades, since their plausible contribution to the diffuse electron background detected at Earth. It might cause anisotropies related to the GMF structure~\cite{jin2020astrophysical,di2021novel,zhu2023prediction}. 

It is important to stress the limitations of the results here reported. The ISRF implemented in the simulations is an approximation of the model: a future work might require the incorporation of a space-dependent photon field in the code framework. We are actively working on implementing this in the CRPropa code. It will be designed for detailed studies of observed gamma-ray sources all around our Galaxy, serving as the natural extension of this work. Besides code concerns, this work assumes specific models for both the ISRF and the GMF. It would be worthwhile to investigate various ISRF and GMF models present in literature, with all the possible updates and scenarios. For instance, a very recent revision of the GMF model, proposing eight coherent component models, is presented in Ref.~\cite{unger2023coherent}. Actually, it would impact on the ratio $B_\text{turb}/B_\text{reg}$, thus on the arrival direction maps. A follow-up interesting work could involve the detailed modelisation of gamma-ray sources or smaller size, i.e. $\lesssim 100 \; \text{pc}$, regions in our Galaxy in the code, e.g. OB super-bubbles whose environments~\cite{chu2007bubbles,joubaud2019gas} are efficient cosmic-ray factories~\cite{binns2011cosmic, nath2020diffuse}.
 
Overall, the significant point would be to compare what learnt from simulations, in terms of gamma-ray absorption and ``deflection", with current and future gamma-ray experiments performance, to properly infer the impact of propagation effects on observed galactic gamma-ray fluxes. This will result in improved characterisations of PeVatrons and galactic centre intrinsic emission models, as well as in more robust constraints from dark matter searches through gamma rays.  


\acknowledgments

This work was funded by the ``la Caixa'' Foundation (ID 100010434) and the European Union's Horizon~2020 research and innovation program under the Marie Skłodowska-Curie grant agreement No~847648, fellowship code LCF/BQ/PI21/11830030. The work of GDM and MASC was also supported by the grants PID2021-125331NB-I00 and CEX2020-001007-S, both funded by MCIN/AEI/10.13039/501100011033 and by ``ERDF A way of making Europe''. The authors also acknowledge the MultiDark Network, ref. RED2022-134411-T. GDM's work is supported by \textit{FPI Severo Ochoa} PRE2022-101820 grant. RAB acknowledges the support from the Agence Nationale de la Recherche (ANR), project  ANR-23-CPJ1-0103-01.

\paragraph*{\textbf{Data availability}} The simulation data performed and employed in this work are accessible for downloading at the data repository of the DAMASCO group website~\cite{repoDamasco}. The files contain also the EM cascade electrons reaching the observer, despite they are not accounted in our analysis. We encourage those interested in the data to reach out the authors for discussions about data structuring and analysis.

\appendix

\section{Magnetic field along the line of sight}\label{AppBfield}

\begin{figure*}
    \centering
    \includegraphics[width=0.93\linewidth]{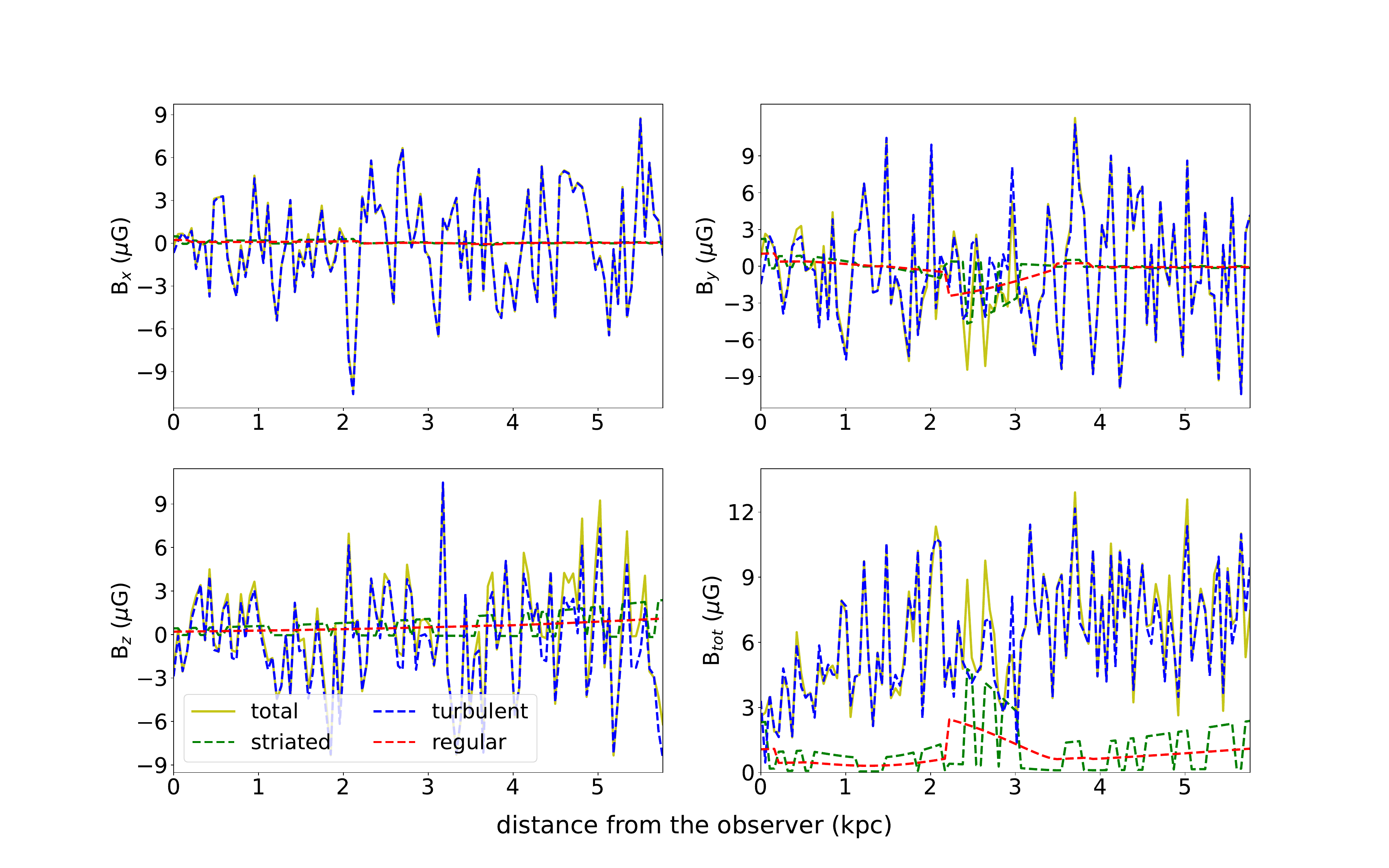}
    \caption{Magnetic field components and total intensity along the line of sight to $S_{<}$~source, $\sim 6 \; \text{kpc}$ distant (see Fig.~\ref{fig:GalBfieldSources}). The GMF model employed is JF12+. The overall magnetic field (\textit{orange line}) is strongly dominated by its turbulent component (\textit{dashed blue line}), so that the former occasionally overlaps with the second.}
    \label{fig:Bfiel_along_los_manysp}
\end{figure*}

\begin{figure*}
    \centering
    \includegraphics[width=0.93\linewidth]{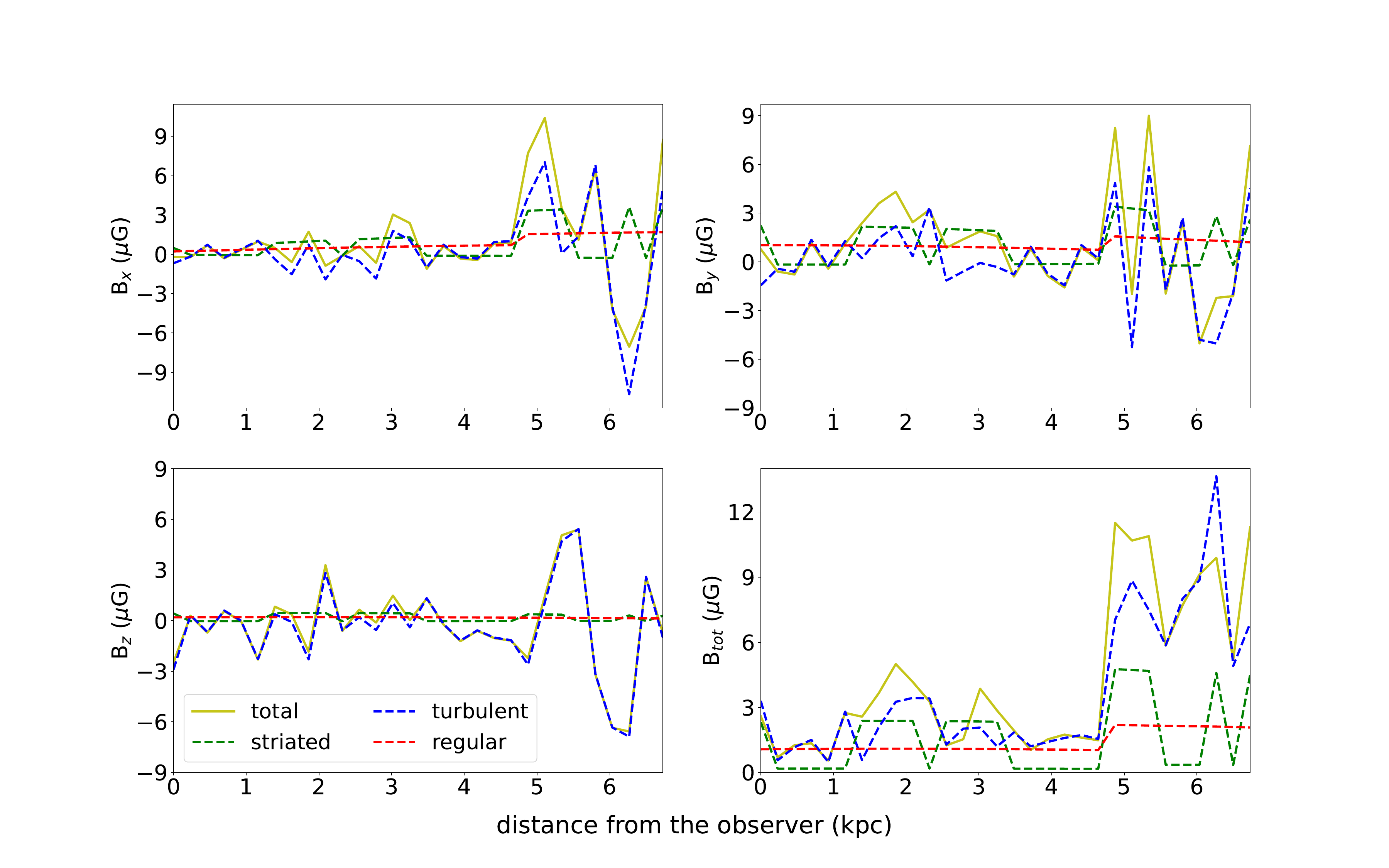}
    \caption{Same as in Fig.~\ref{fig:Bfiel_along_los_manysp}, but in this case the GMF structure refers to the line of sight towards the $S_{1}$~source (see Fig.~\ref{fig:GalBfieldSources}).}
    \label{fig:Bfiel_along_los_spiral}
\end{figure*}

The GMF structure in JF12+ comprehends the regular, striated, and turbulent components, as explained in Sec.~\ref{sub_magnModl}. In Fig.~\ref{fig:Bfiel_along_los_manysp} and Fig.~\ref{fig:Bfiel_along_los_spiral}, two GMF configurations are depicted along two lines of sight, namely the one related to the $S_{<}$~source, $\sim 6 \; \text{kpc}$ distant, and the other to $S_{1}$~source. While in the former the line of sight goes through many GMF spirals, the latter crosses only one spiral. This fact is reflected in the interplay between regular and turbulent components. In the ratios we neglect the contribution from the random striated field since it is expected to be negligible in most of the accounted galactic regions, according to JF12+.

\section{Observer size}\label{AppObsSize}

\begin{figure}
    \centering
    \includegraphics[width=0.49\linewidth]{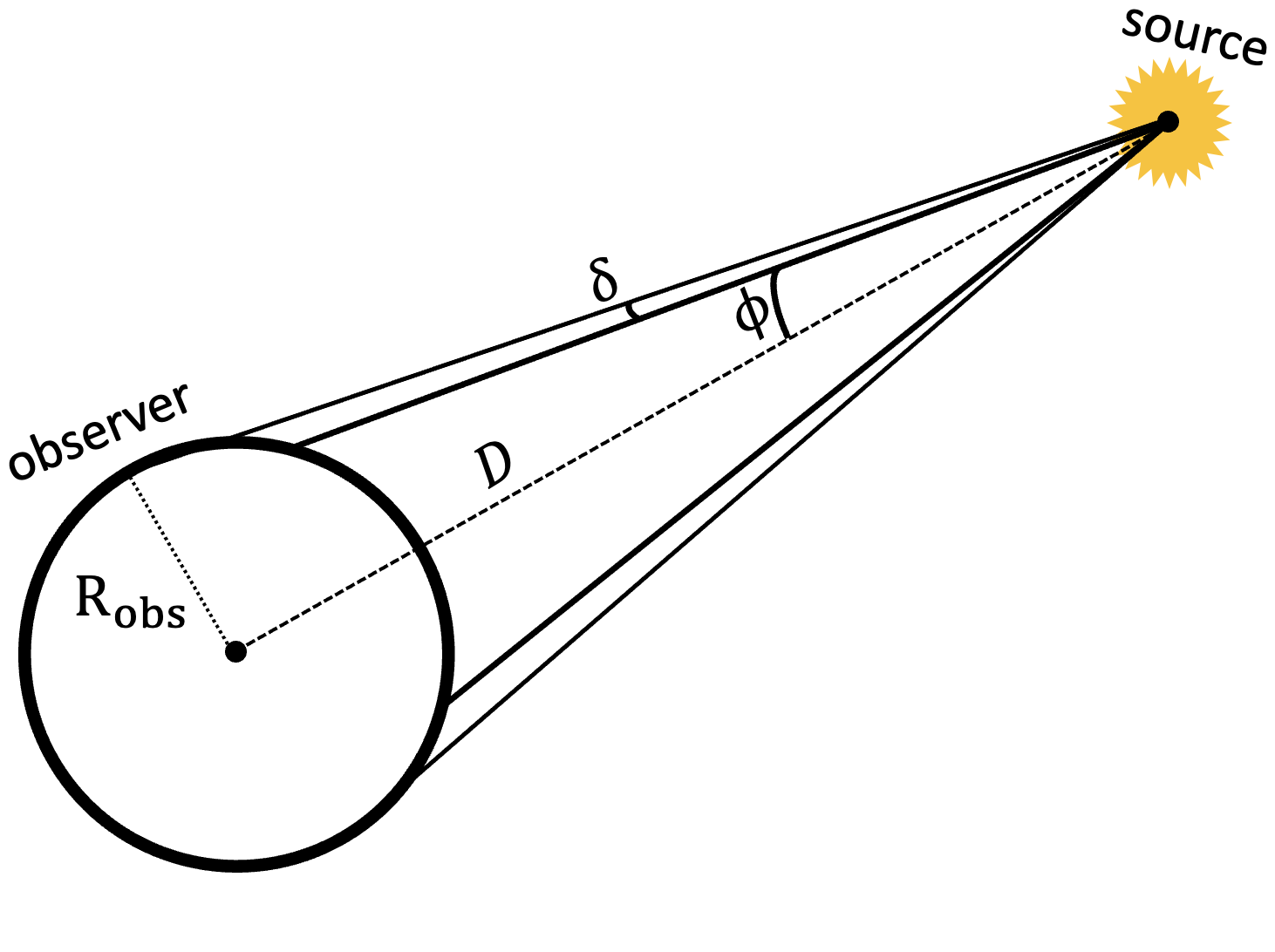}
    \caption{Not-to-scale diagram of the observer radius geometry.}
    \label{fig:sketchObs}
\end{figure}

The observer of the simulations, positioned at the nominal Earth position, is spherical in shape. In our work, the sphere size is geometrically related to the angular aperture of the intrinsic emission cone of the source, given that the jet axis points towards the centre of the observer sphere. The geometrical radius is $R = D\cdot \tan\phi$, with $\phi$ denoting half-aperture of the prompt emission cone and $D$ the observer-source distance. However, we add a tiny angle $\delta$ to the half-aperture $\phi$ since, if a gamma ray originates with an initial direction in the outer layers of the cone, the observer may detect minute deflections. Therefore the observer radii chosen in our simulations are $R_\text{obs} = D\cdot \tan(\phi + \delta)$, as depicted in Fig.~\ref{fig:sketchObs}.

In general, the purpose in choosing a certain radius is to find a compromise between the smallest observer size and statistics, thus the amount of detected gamma rays. The latter is directly related to computational time. To avoid altering the results by systematic trends due to a specific choice, it is important to evaluate two relevant quantities: the electrons mean free paths and the typical length scales of the curvature of magnetic field lines. 
Both quantities are due to the environmental properties in the surroundings of the observer. The former is the inverse of Eq.~\ref{InvMFP} in the case of IC scattering, reported in Fig.~\ref{fig:elEnergyLosses}, thus the electron mean free path. The latter is interpreted as the magnetic field curvature length scale; it is expressed as $\big(\frac{\Vec{\nabla}\times\Vec{B}}{\Vec{B}}\big)^{-1}$, where $\Vec{B}$ is the magnetic field. Simulations, that include more accurate and detailed ISRF and GMF models in the Earth surroundings, will need to take into account the two lengths, just introduced. In this case, to reduce the impact of systematics on simulation results, the observer radius will need to be chosen as the minimum between the electron interaction mean distances and the magnetic field curvature scale.

\begin{figure*}
    \centering
    \includegraphics[width=0.6\textwidth]{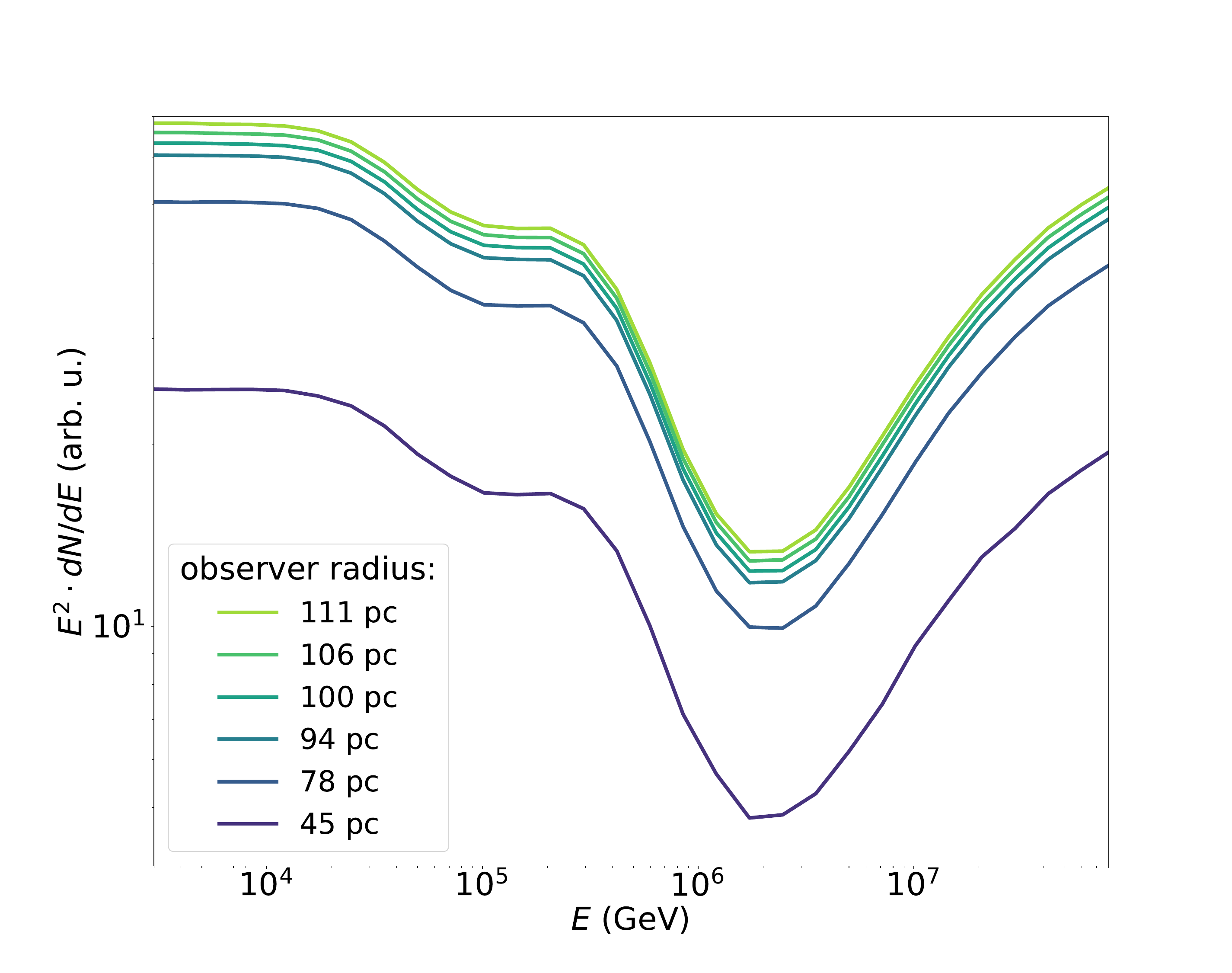}
    \caption{Energy spectra from the $S_{>}$~source around 11~kpc far from the spherical observers. The spheres with various radii are centered at Earth position.}
    \label{fig:energySpectraManyObs}
\end{figure*}

In Fig.~\ref{fig:energySpectraManyObs}, we show the observed energy spectra in dependence to the spherical observer radius. It ranges from the geometrical one ($\sim$~110~pc) to the average magnetic field curvature length near the Earth ($\sim$~45~pc) for the JF12+ model. The energy spectra features, discussed in Sec.~\ref{Spectra}, are similar for the six observer radii set in the simulation.

\section{Arrival direction maps}\label{AppCounntMaps}
\begin{figure*}

     \begin{minipage}[t]{0.48\textwidth}
         \centering
         \includegraphics[width=\linewidth]{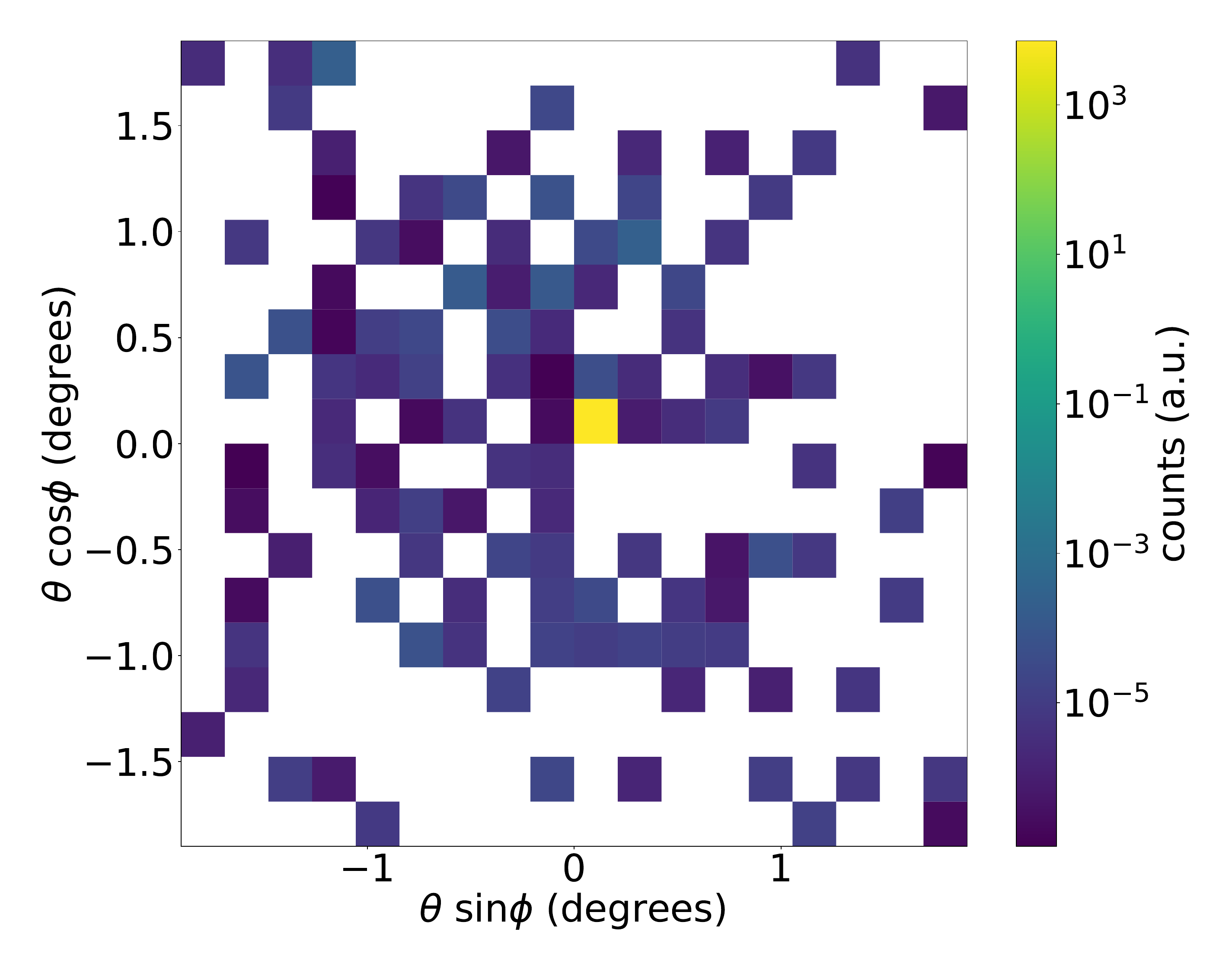}
        \captionsetup{labelformat=empty}    
         \caption*{(a) Source coordinates: $x=-3$~kpc, $y=1.9$~kpc. The distance source-observer is $\sim 6 \; \text{kpc}$.}
    \end{minipage}
    \hfill
    \begin{minipage}[t]{0.48\textwidth}
         \centering
         \includegraphics[width=\linewidth]{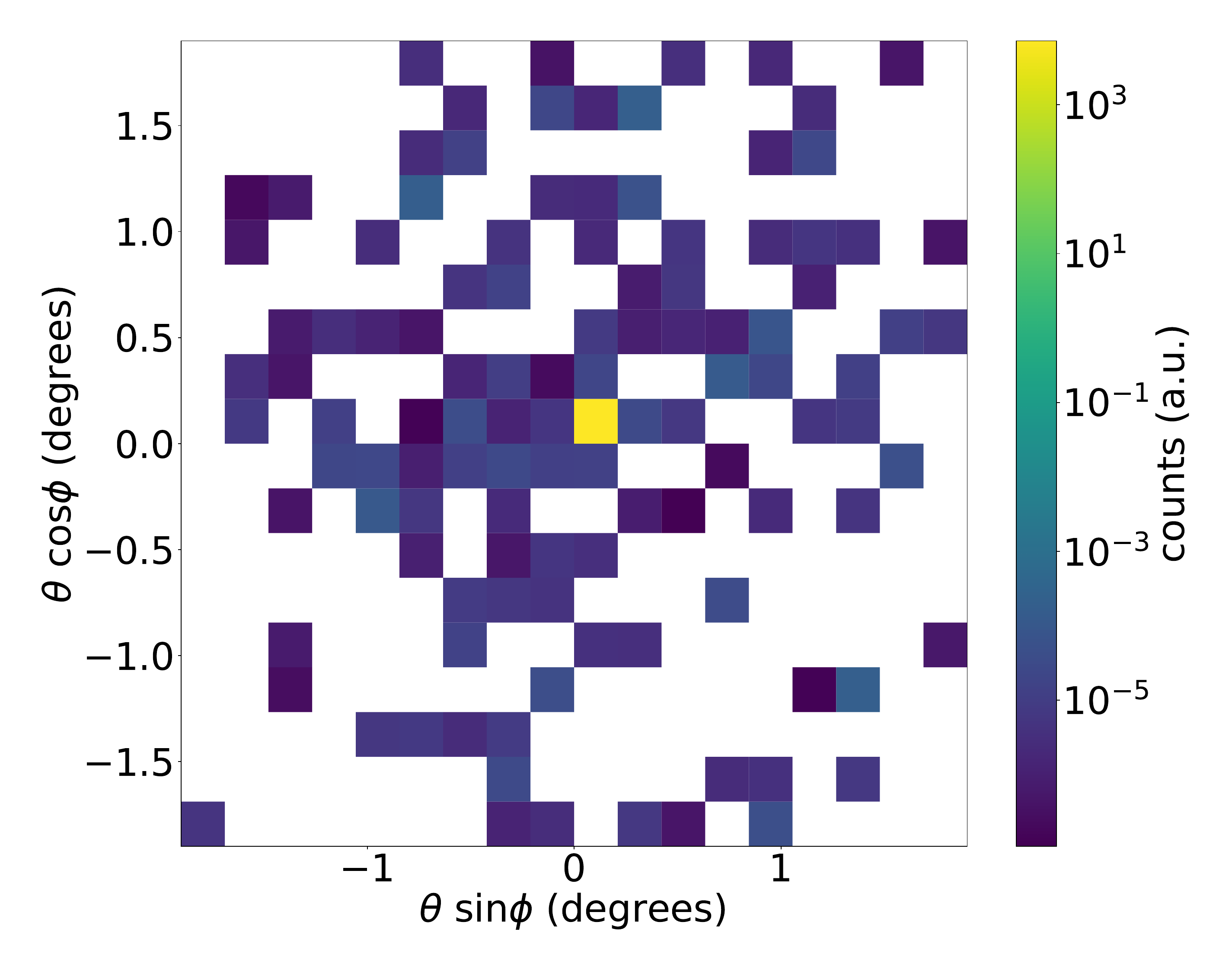}
         \captionsetup{labelformat=empty}
         \caption*{(b) Coordinates $x=-3$~kpc, $y=-1.9$~kpc.}
     \end{minipage}
     \hfill
     \begin{minipage}[t]{0.48\textwidth}
         \centering
         \includegraphics[width=\linewidth]{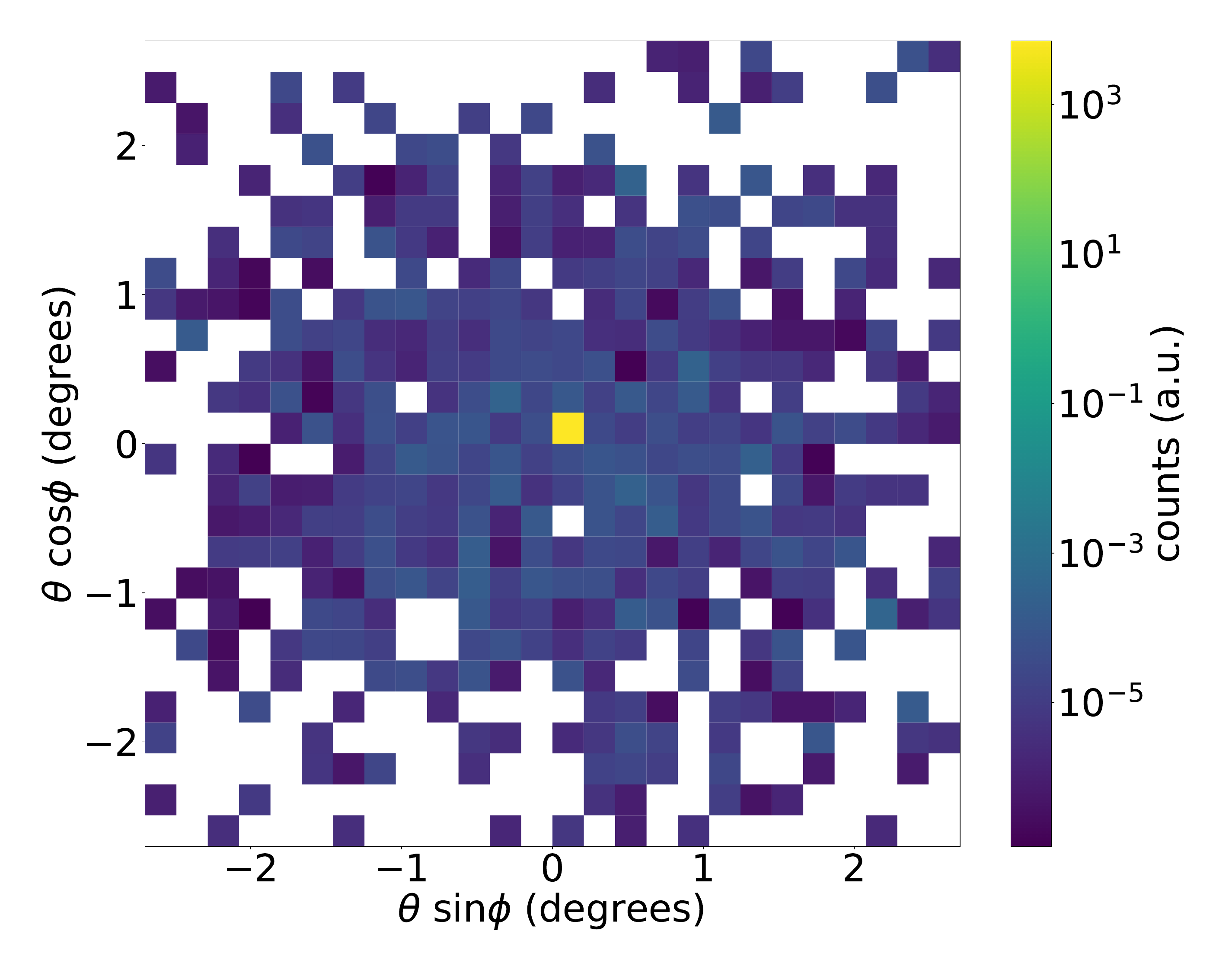}
        \captionsetup{labelformat=empty}
        \caption*{(c) Coordinates $x=2$~kpc, $y=3.7$~kpc. The distance source-observer is $\sim 11 \; \text{kpc}$. }
    \end{minipage}
    \begin{minipage}[t]{0.48\textwidth}
         \centering
         \includegraphics[width=\linewidth]{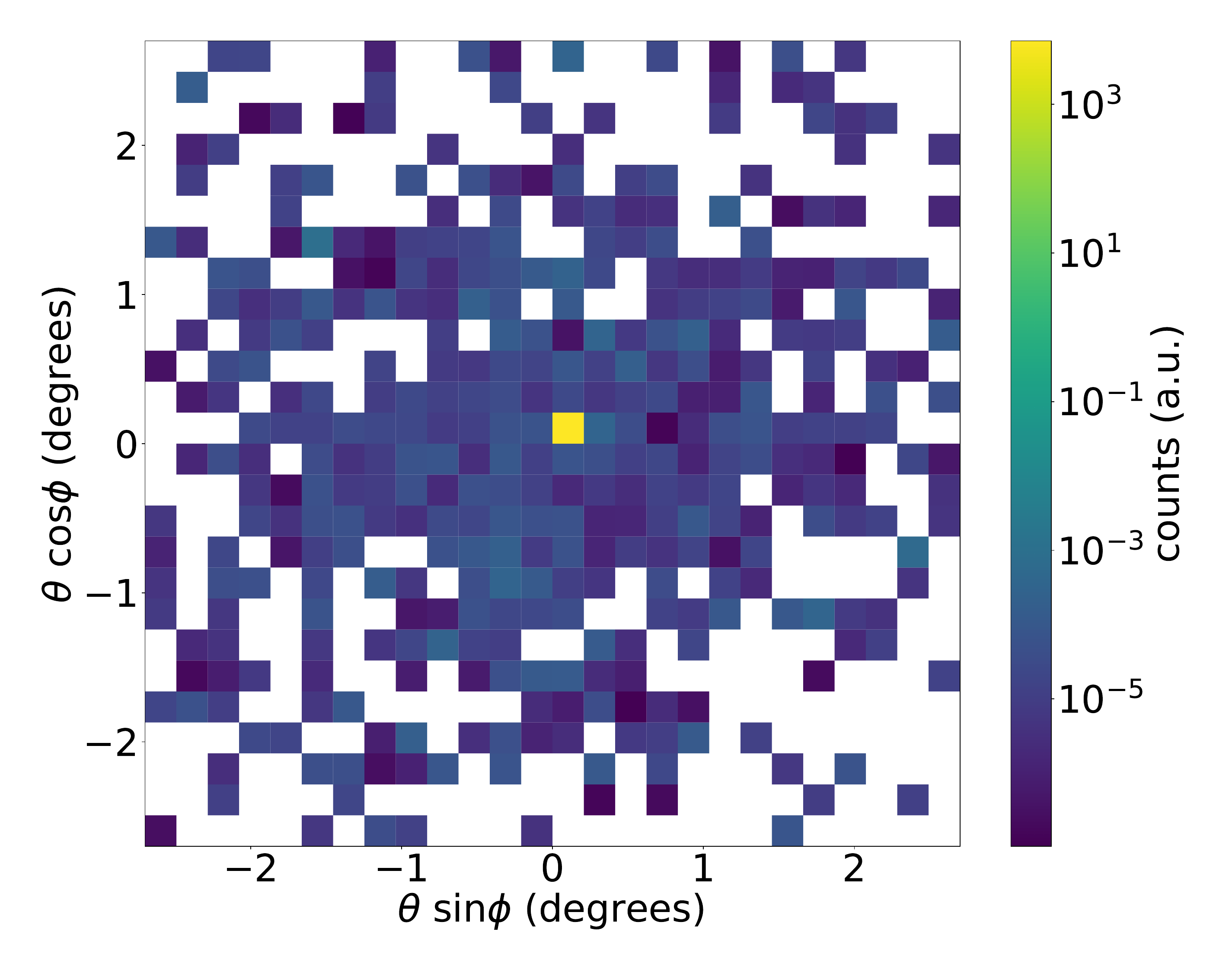}
         \captionsetup{labelformat=empty}
         \caption*{(d) Coordinates $x=2$~kpc, $y=-3.7$~kpc.}
         \captionsetup{labelformat=empty}
     \end{minipage}
     \caption{Count maps of the closest $S_{>}$~\textit{(left)} and $S_{<}$~\textit{(right)} set of sources. The $(x,y)$ coordinates refer to the cartesian position in our Galaxy, as shown in Fig.~\ref{fig:GalBfieldSources}.}
     \label{fig:map_plots_close}
\end{figure*}

\begin{figure*}
    
    \begin{minipage}[t]{0.48\textwidth}
         \centering
         \includegraphics[width=\linewidth]{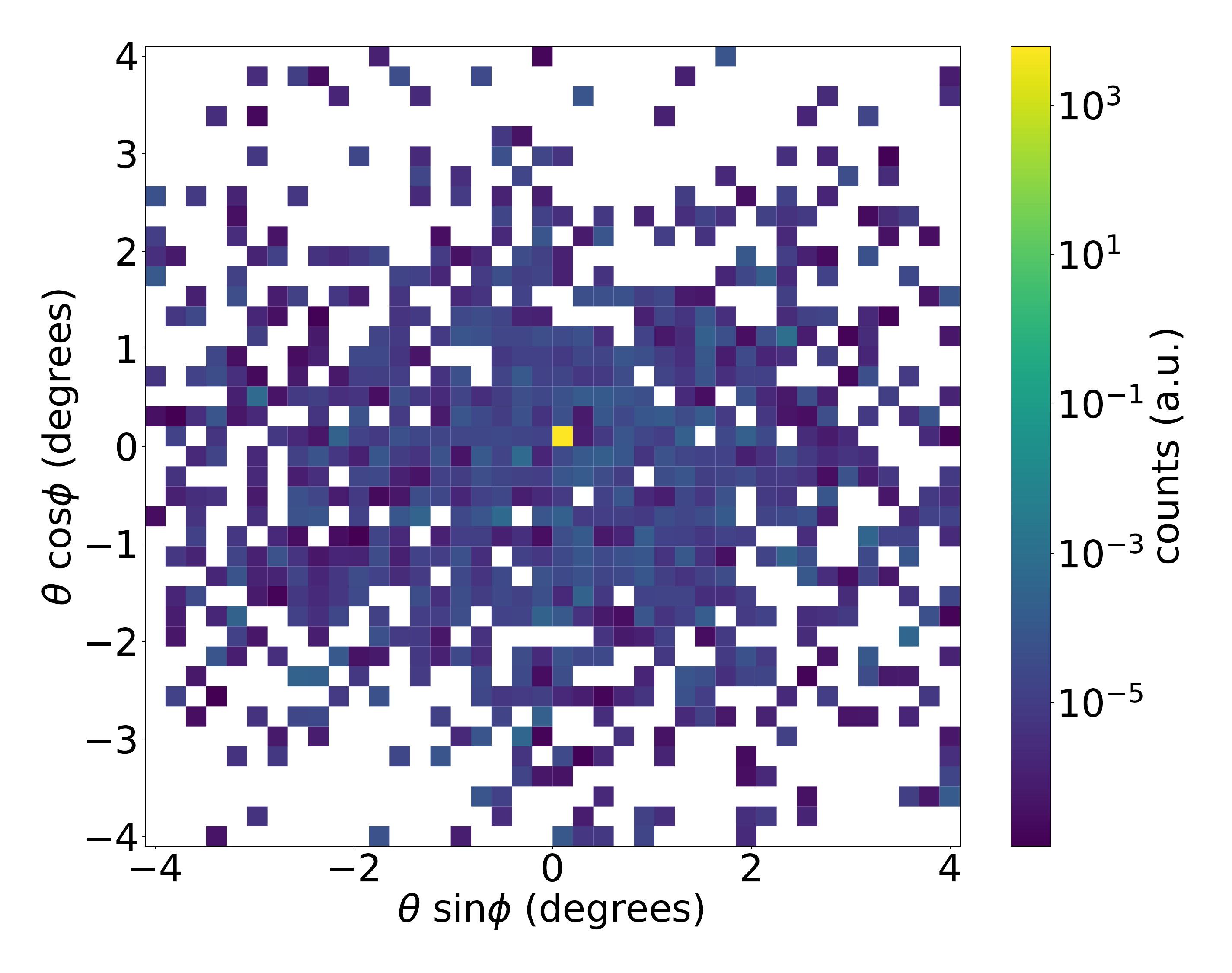}
         \captionsetup{labelformat=empty}
         \caption*{(e) Coordinates $x=8.5$~kpc, $y=6$~kpc. The distance source-observer is $\sim 18 \; \text{kpc}$.}
    \end{minipage}
    \hfill
    \begin{minipage}[t]{0.48\textwidth}
         \centering
         \includegraphics[width=\linewidth]{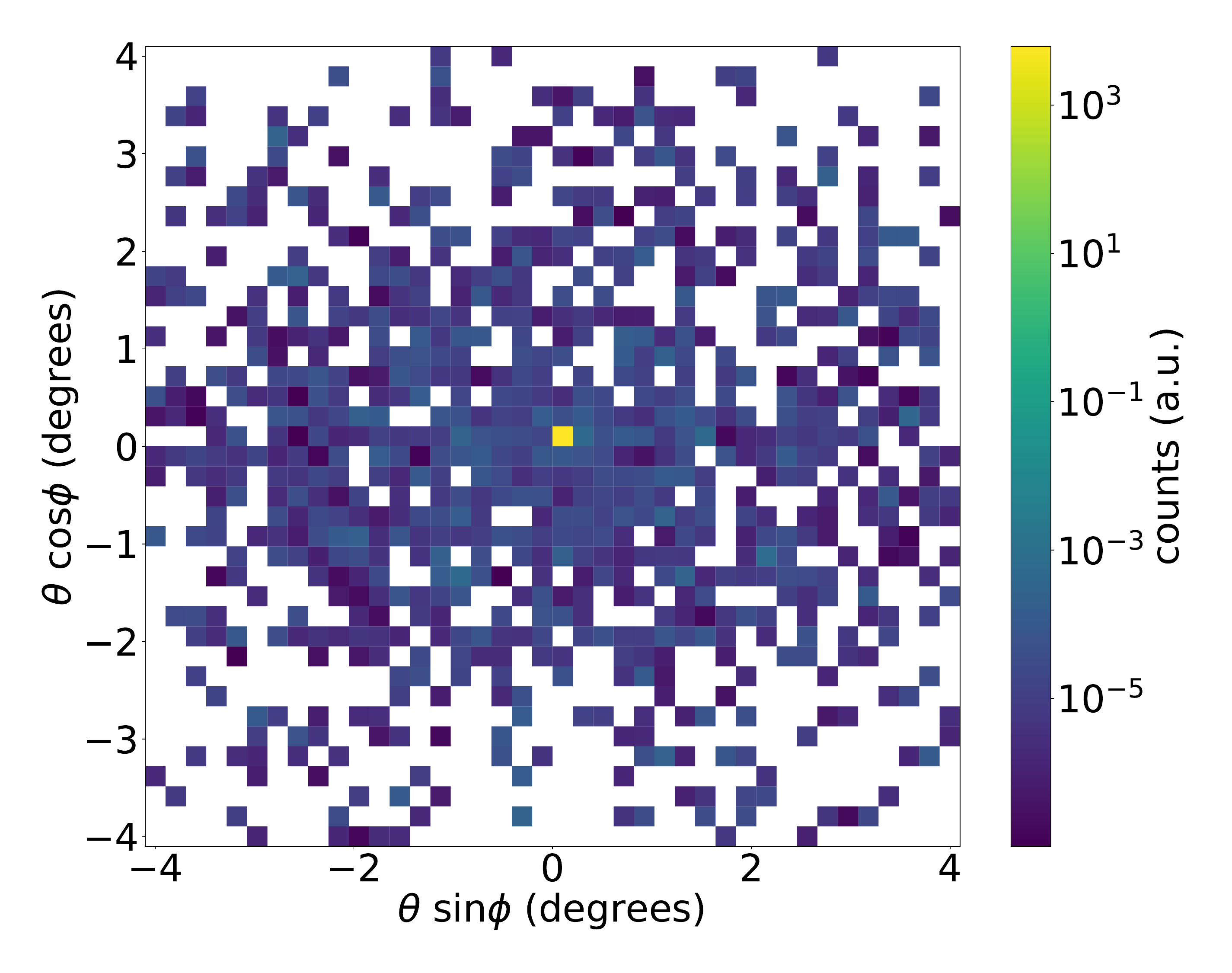}
         \captionsetup{labelformat=empty}
         \caption*{(f) Coordinates $x=8.5$~kpc, $y=-6$~kpc.}
    \end{minipage}
    \hfill
    \begin{minipage}[t]{0.48\textwidth}
         \centering
         \includegraphics[width=\linewidth]{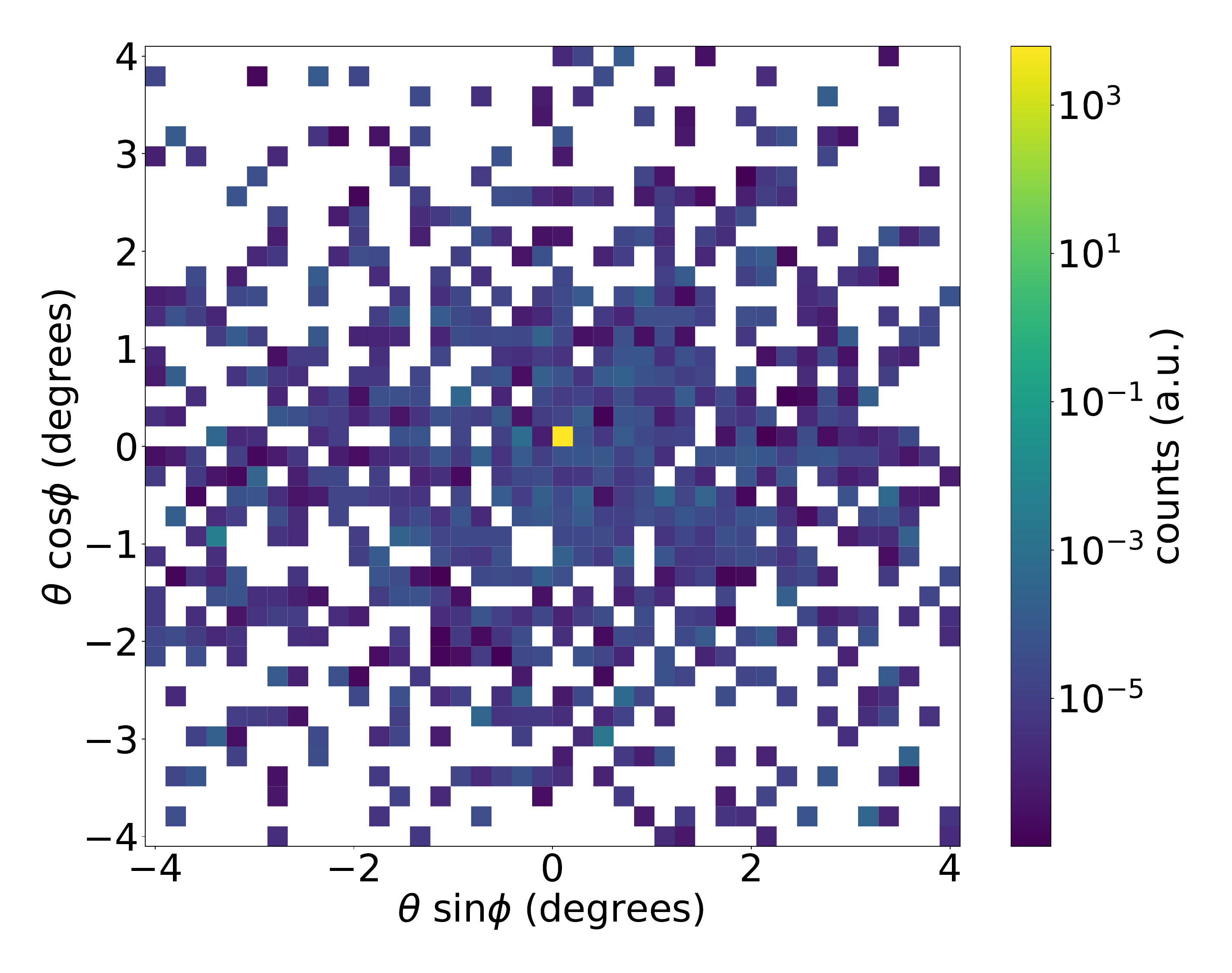}
         \captionsetup{labelformat=empty}
         \caption*{(g) Coordinates $x=15$~kpc, $y=8.3$~kpc. The distance source-observer is $\sim 25 \; \text{kpc}$.}
    \end{minipage}
    \hfill 
    \begin{minipage}[t]{0.48\textwidth}
         \centering
         \includegraphics[width=\linewidth]{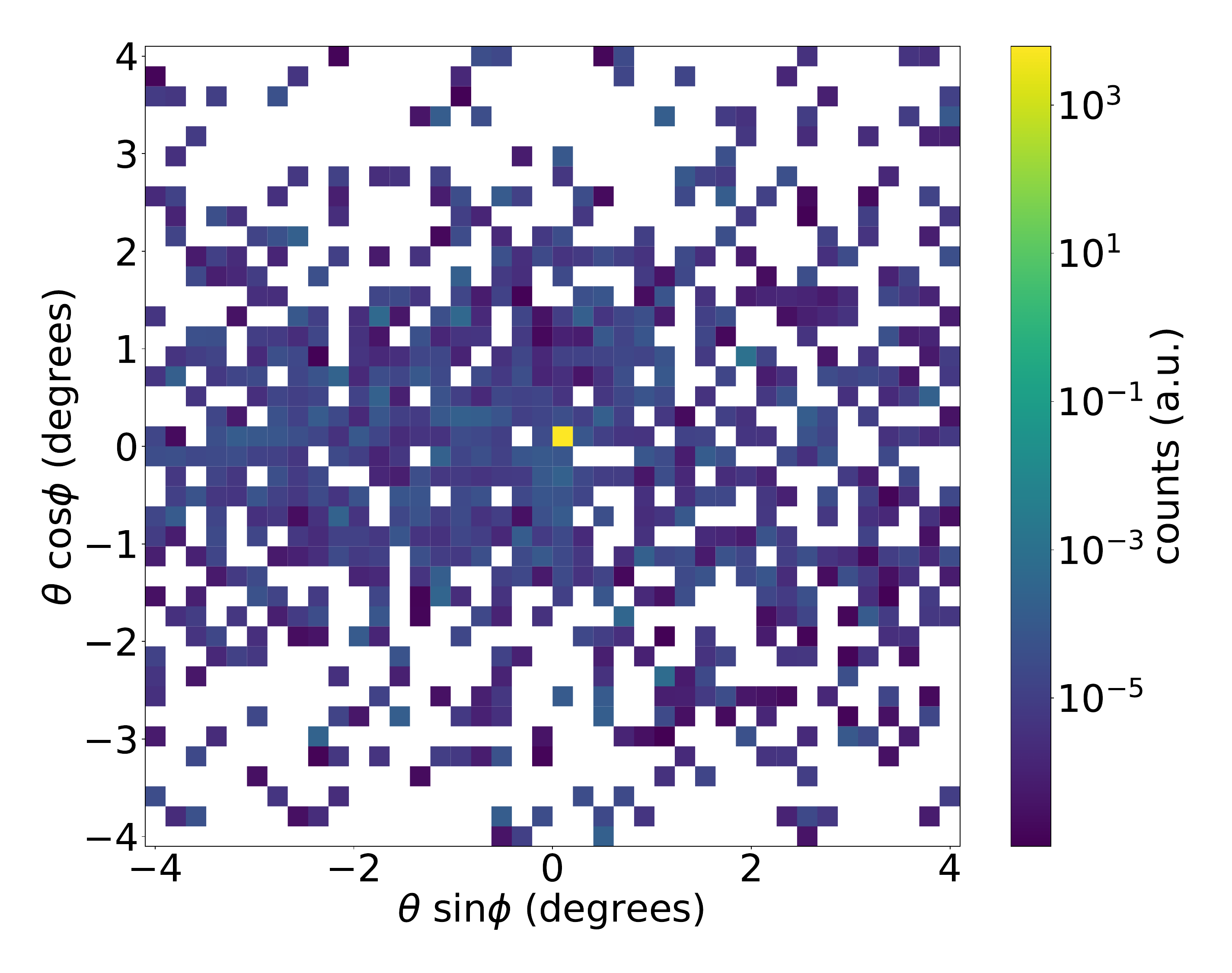}
         \captionsetup{labelformat=empty}
         \caption*{(h) Coordinates $x=15$~kpc, $y=-8.3$~kpc.}
     \end{minipage}
\captionsetup{list=no}
\caption{Same as in Fig.~\ref{fig:map_plots_close} but for the farthest $S_{>}$~\textit{(left)} and $S_{<}$~\textit{(right)} set of sources.}
\label{fig:map_plots_far}
\end{figure*}

\begin{figure*}

    \begin{minipage}[t]{0.48\textwidth}
        \centering
        \includegraphics[width=\linewidth]{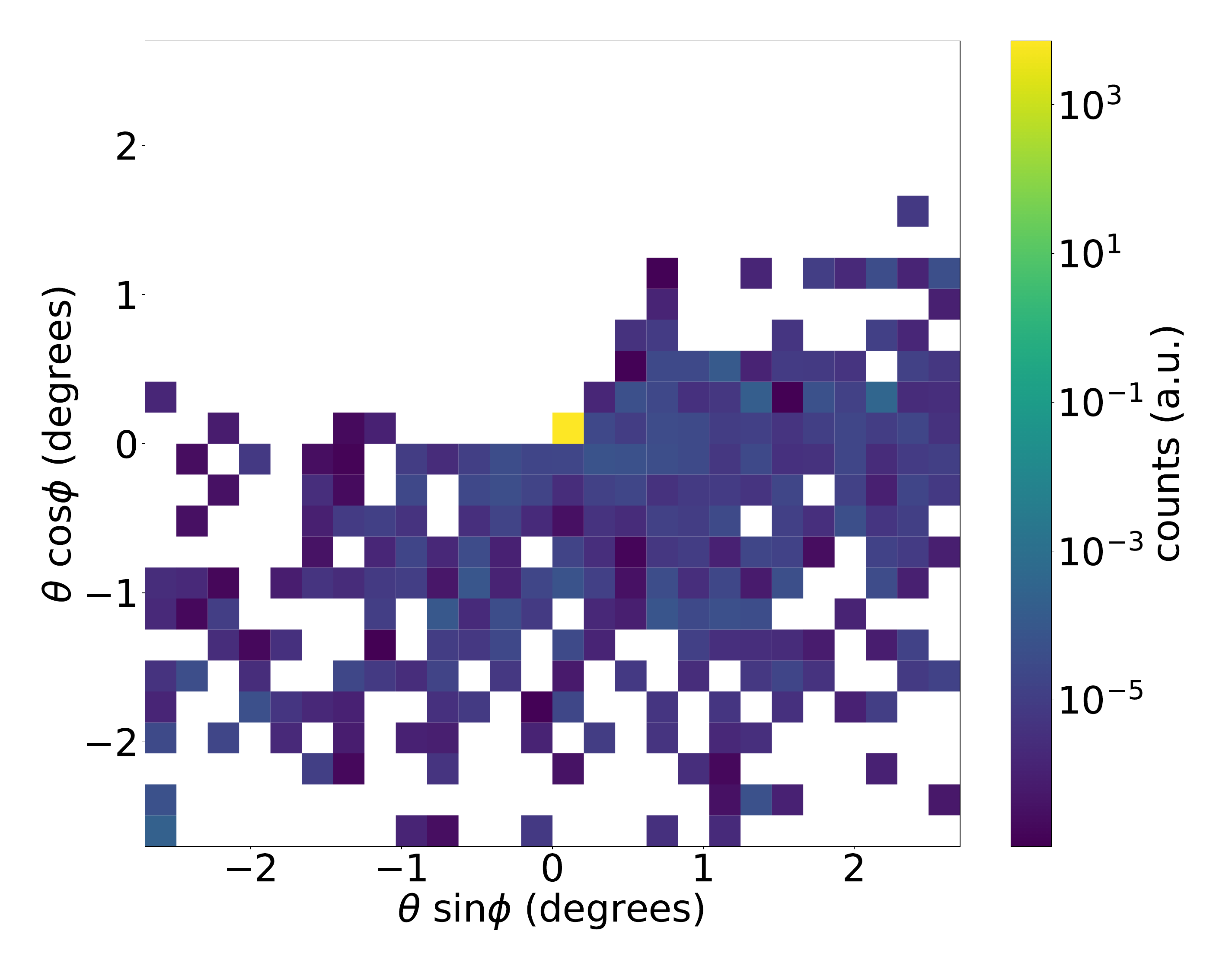}
        \caption{Source coordinates $x=-7$~kpc, $y=6.8$~kpc, named $S_{1}$ in Fig.~\ref{fig:GalBfieldSources}. The distance source-observer is $\sim 7 \; \text{kpc}$.}
        \label{fig:PP0_7kpc}
    \end{minipage}
    \hfill
    \begin{minipage}[t]{0.48\textwidth}
         \centering
         \includegraphics[width=\linewidth]{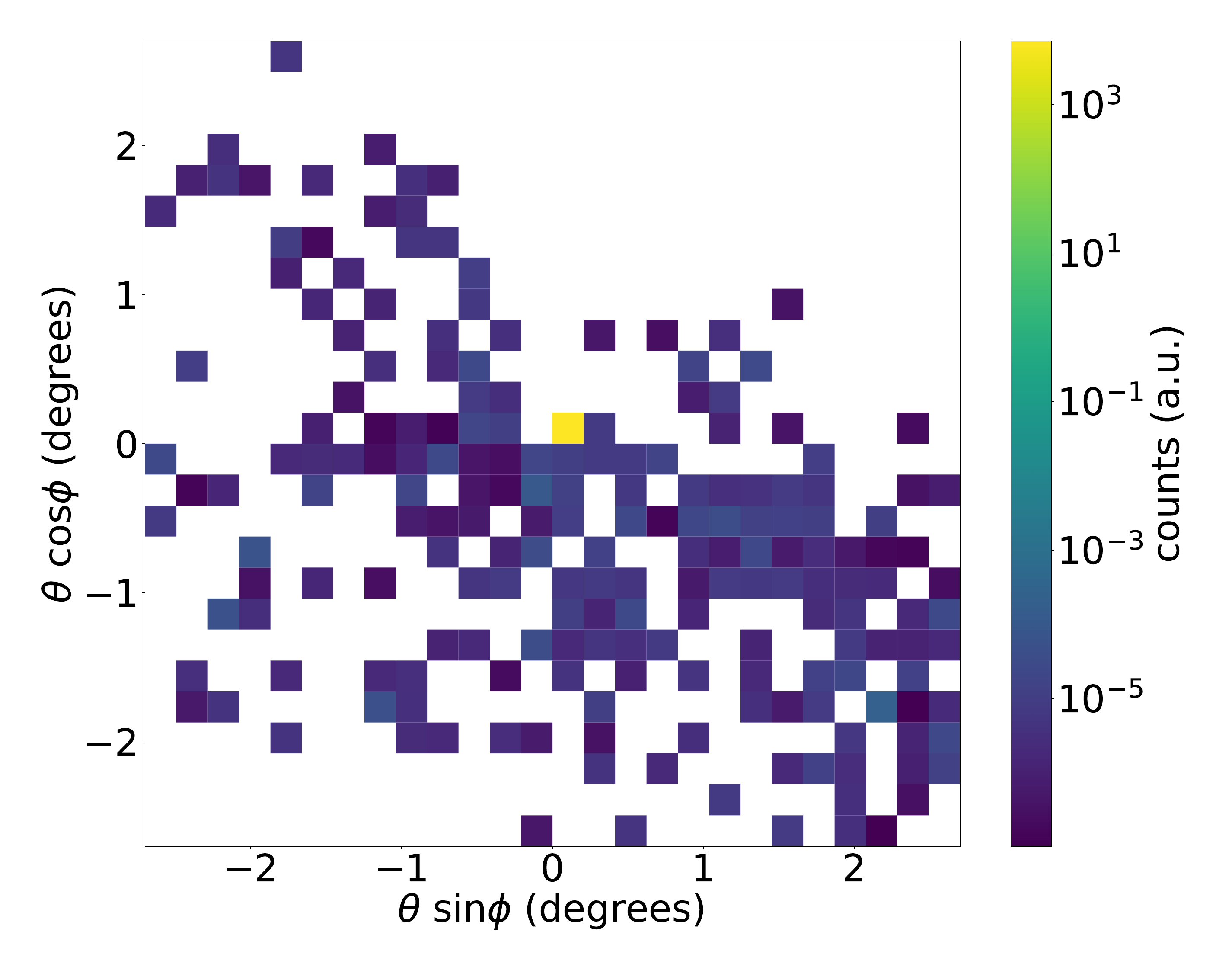}
         \caption{Coordinates $x=-6$~kpc, $y=5$~kpc, named $S_{2}$ source. The distance source-observer is $\sim 5.6 \; \text{kpc}$.}
         \label{fig:PP0_5.6kpc}
    \end{minipage}
\end{figure*}

This appendix contains the spatial count maps for the ten sources simulated in this work, as discussed in Sec.~\ref{haloes}. The count maps of $S_{\lessgtr}$ sources are in Fig.~\ref{fig:map_plots_close} and Fig.~\ref{fig:map_plots_far}, respectively for the closest and farthest sources. The ones corresponding to $S_{1}$ and $S_{2}$ are shown in Fig.~\ref{fig:PP0_7kpc} and Fig.~\ref{fig:PP0_5.6kpc}, respectively. The amount of injected gamma rays in the simulation of each source is $10^{7}$. To note that the values on the \textit{colorbars} are not actual counts since, in our simulations, the \textit{thinning} procedure is applied to sample the secondary particles in the cascade~\cite{batista2022crpropa}.  

\bibliography{main}

\end{document}